%% file: MEM_ptp.tex
\documentclass[seceq]{ptptex}
\newcommand{\siml}{\hspace{.3em}\raisebox{.4ex}{$<$}\hspace{-.75em}
 \raisebox{-.7ex}{$\sim$}\hspace{.3em}}
\newcommand{\simg}{\hspace{.3em}\raisebox{.4ex}{$>$}\hspace{-.75em}
 \raisebox{-.7ex}{$\sim$}\hspace{.3em}}
\usepackage{graphicx}
\usepackage{bm}

\preprintnumber[4.5cm]{
SAGA-HE-202\\YAMAGATA-HEP-03-30\\ September, 2003}

\title{
Maximum Entropy Method Approach to the  $\theta$ Term
}

\author{
 Masahiro Imachi$^{\ddagger}$\footnote{E-mail: imachi@sci.kj.yamagata-u.ac.jp},\\
  Yasuhiko Shinno$^{\diamond}$\footnote{E-mail: 
  shinno@dirac.phys.saga-u.ac.jp} and 
   Hiroshi Yoneyama$^{\diamond}$\footnote{E-mail: yoneyama@cc.saga-u.ac.jp}
}

\inst{
 Department of Physics, Yamagata University, Yamagata 990-8560, Japan$^{\ddagger}$\\
 Department of Physics, Saga University, Saga 840-8502, Japan$^{\diamond}$\\}

\abst{
In Monte Carlo simulations of  lattice field theory with a 
$\theta$ term, one confronts the complex weight problem, or the sign 
problem. This is circumvented  by performing  the Fourier transform of 
the topological charge distribution $P(Q)$. This procedure, however, 
  causes  flattening phenomenon of the free energy $f(\theta)$, 
  which makes study of the  phase structure unfeasible. 
   In order to treat this problem,  
  we apply the maximum 
 entropy method (MEM) to a  Gaussian form of  $P(Q)$, which serves as a good 
 example to test whether the MEM can be applied  effectively to  the $\theta$ term. 
 We study the case   with   flattening as well as that  without   
 flattening. In the latter case, the results of the MEM agree with 
 those obtained from the direct
  application of the  Fourier transform.   For the former,   the MEM gives 
  a  smoother  $f(\theta)$  than that of the  Fourier transform.  Among 
   various default models investigated,  the images which  yield the least 
  error do not  show flattening, although some others  cannot be 
   excluded given the  uncertainty related to statistical error.    
}

\begin{document}

\maketitle

\input{sec1}
\input{sec2}
\input{sec3}
\input{sec4}
\input{sec5}
\section*{Acknowledgements}
We are grateful to Professor  Shoichi Sasaki and Professor  Masayuki Asakawa 
for useful discussions
about the MEM.
This work is supported in part by  Grants-in-Aid for  Scientific Research (C)(2) of 
the  Japan Society for the Promotion of Science  (No. 15540249)
and of the Ministry of Education, Culture, Sports, Science and 
Technology (Nos. 13135213 and 13135217).
 Numerical calculations were performed  partly  on a computer 
at the  Computer and Network  Center, Saga University.
%
\appendix
\input{appendix1}
\input{appendix2}
\input{appendix3}
\input{appendix4}

\end{document}

%% file: sec1.tex
\section{Introduction}
\label{sec:introduction}
\par
  It is believed  that the $\theta$ term could affect the
  dynamics at low energy and the vacuum structure of QCD, but it is known
  from experimental evidence that the value of $\theta$ is strongly 
  suppressed in Nature. From the theoretical point of view, the reason 
  for this is not clear yet. Hence, it is important to study the properties 
  of QCD with the
  $\theta$ term to clarify the structure of the QCD 
  vacuum.~\cite{rf:tHooft} \  For  theories with the $\theta$ term, 
  it has been  pointed out that  rich phase
  structures could be realized in
  $\theta$ space. For example, the phase structure
  of the  $Z(N)$ gauge model was investigated using free energy arguments,  and
  it was found that
  oblique confinement phases could occur.~\cite{rf:CR}  \ 
  In CP$^{N-1}$ models, which have
  several dynamical properties in common with QCD, it has been  shown that a
  first-order phase transition exists at
   $\theta=\pi$~\cite{rf:Seiberg,rf:HITY,rf:IKY}. \par
  Although  numerical simulation is one of the most effective tools to study
  non-perturbative properties of field theories, the introduction of the
  $\theta$ term makes the Boltzmann weight complex. This makes it
  difficult to perform  Monte Carlo (MC) simulations on a
  Euclidean lattice. This is the complex action problem, or the sign
  problem. In order to circumvent this problem, the following method is
  conventionally employed.~\cite{rf:BRSW,rf:Wiese}  \  The partition function 
${\cal Z}(\theta)$ can
  be obtained by Fourier-transforming the topological charge distribution
  $P(Q)$, which is calculated with a  real positive Boltzmann weight:
\begin{equation}
  {\cal Z}(\theta)=\frac{\int[d\bar{z}dz]e^{-S+i\theta\hat{Q}(\bar{z},z)}}
   {\int[d\bar{z}dz]e^{-S}}\equiv\sum_{Q}e^{i\theta Q}P(Q),
    \label{eqn:partitionfunction}
\end{equation}
  where
\begin{equation}
  P(Q)\equiv \frac{\int[d\bar{z}dz]_Qe^{-S}}{\int d\bar{z}dz e^{-S}}.
   \label{eqn:Pq}
\end{equation}
  The measure $[d\bar{z}dz]_Q$ in Eq.~(\ref{eqn:Pq}) is such  that the 
integral is
  restricted to configurations of the field $z$ with  topological 
charge $Q$. Also, 
  $S$ represents the  action. \par
  In the study  of CP$^{N-1}$ models, it is known that this algorithm
  works well for a small lattice volume $V$ and in the strong coupling
  region~\cite{rf:BRSW,rf:HITY,rf:PS,rf:BISY}. \ 
  As the volume increases or in the weak coupling region,  however,
  this strategy too suffers from the
  sign problem for $\theta\simeq\pi$. The error in 
  $P(Q)$ masks the true values of ${\cal Z}(\theta)$ in the vicinity of
  $\theta=\pi$, and this results in  a fictitious signal of a phase
  transition~\cite{rf:OS}. \  This is called `flattening', because the free
  energy becomes almost flat for $\theta$ larger than a certain value.
  This problem  could be remedied by reducing the error in  $P(Q)$. 
  This, however, is hopeless,  because  the  
  amount of data needed to reduce the error to a given level 
  increases exponentially with $V$.
   Recently, an alternative method has
  been   proposed to circumvent  the sign problem.~\cite{rf:ACGL,rf:AANV}
  \par
  Our aim  in the present
  paper is to reconsider this problem  from the point of view of the maximum
  entropy method (MEM)~\cite{rf:Bryan,rf:SSG,rf:GJSS,rf:JGU,rf:AHN}. \  The
   MEM is well known as a powerful tool for
  so-called  ill-posed problems, where the number of parameters to be
  determined is much larger than the number of data points. It has 
  been  applied to
  a wide range of fields,  such as radio astrophysics and
  condensed matter physics. Recently,  spectral
  functions in lattice field theory  have been widely studied  by use of
  the  MEM~\cite{rf:AHN,rf:Yamazaki_CP-PACS,rf:Sasaki,rf:YI}. \ 
   In the present paper, we are interested in whether the MEM can be 
   applied effectively to 
   the study of the $\theta$ term and to what extent one can improve
  the flattening phenomenon of the free energy. \par
  The MEM is based upon  Bayes' theorem. It derives the most probable
  parameters by utilizing data sets and our knowledge about these
  parameters in terms of the probability. The probability distribution,
  which is called the posterior probability, is given by the product of the
  likelihood function and the prior probability. The latter is
  represented by the Shannon-Jaynes entropy, which plays an important
  role to guarantee the uniqueness of the solution, and the former is
  given by $\chi^2$. It should be noted that artificial
  assumptions are not needed in  the calculations, because
  the determination of a unique solution is carried out  according to 
  probability theory.
   Our task is to determine  the image for which  the
  posterior probability is maximized. In practice, however,
  it is difficult to find a unique solution in  the huge configuration space
  of the image. In order to do so effectively, we employ the singular
  value decomposition (SVD). \par
  The flattening of the free energy is an inherent phenomenon in  the
  Fourier transformation procedure. It is quite independent of the 
  models used.  We choose a Gaussian form of $P(Q)$, which
  is  realized in many  cases,   such as  the strong coupling region
   of the CP$^{N-1}$ model and the
  2-d U(1) gauge model. Because the Gaussian $P(Q)$ can be 
  analytically Fourier-transformed to ${\cal Z}(\theta)$, it provides  a good
  example to investigate whether the  MEM would be effective. For the
  analysis, we use mock data by adding noise to $P(Q)$ in the cases
  with and without flattening. The most 
  probable images of the partition function are obtained. The 
  uncertainty of the images is used as an estimate of  the error.\par
  Our conclusion is summarized as  follows.
\vspace*{4mm}
\begin{enumerate}
  \item In the case without  flattening, the results of the MEM are 
  consistent  with
        those  of the Fourier transformation and thus reproduce the exact
        results.
  \item In the case with  flattening, the MEM yields a  
  smoother free energy density    than  the  Fourier transform.  Among 
   various default models investigated, some images with the  least 
  errors do not exhibit   flattening. 
   With regard  to the question of which is the best 
  among such  images,  further consideration of the systematic error
  is  needed to   check   the robustness of the images.
  \end{enumerate}
\vspace*{4mm}

  This paper is organized as follows. In the following section, we 
  give an 
  overview of the origin of flattening. In $\S$~\ref{sec:MEM}, 
  we
  summarize the procedure for the analysis of the MEM. The results obtained 
by use of
  the MEM are presented in $\S$~\ref{sec:result}. A summary is given  in
  $\S$~\ref{sec:summary}. \par

%% file: sec2.tex
\section{Sign problem and flattening behavior of the  free energy}
\label{sec:signproblem}\par

    In this section, we briefly review the flattening phenomenon of the free
   energy density. It  is observed when one employs
   an  algorithm in  which ${\cal Z}(\theta)$ is calculated through the
   Fourier transform.
   In order to obtain ${\cal Z}(\theta)$, we must
   calculate $P(Q)$ with  high precision. Although
   $P(Q)$ is calculated over a wide range of orders  by use of the set
   method~\cite{rf:KSC}, \  the error in  $P(Q)$ yields error in  ${\cal
   Z}(\theta)$ through the Fourier transform. This  effect becomes serious
   in the  large $\theta$ region.  Here, we
   use a  Gaussian $P(Q)$ for our investigation.
   The Gaussian $P(Q)$ is not just a toy model,  but indeed it 
   is realized  in a variety of models,  such as the 2-d U(1) gauge model and 
in the
   strong coupling limit of CP$^{N-1}$ models. \par
   We parameterize the Gaussian $P(Q)$ as 
\begin{equation}
   P(Q)=A \exp[-\frac{c}{V}Q^2], \label{eqn:Pqparametrize}
\end{equation}
   where, in the case of the 2-d U(1) gauge model, $c$ is a constant depending
   on the inverse coupling constant $\beta$,  and $V$ is the lattice
   volume. Hereafter,  $V$ is regarded as a parameter and varied in the
analysis.
   The  constant  $A$ is  fixed  so that   $\sum_Q P(Q)=1$.
   The distribution $P(Q)$ is analytically transformed by use of the 
Poisson sum
   formula into the partition function
\begin{equation}
   {\cal Z}_{\rm pois}(\theta)=A\sqrt{\frac{\pi V}{c}}\sum_{n=-\infty}^{\infty}
    \exp\biggl[-\frac{V}{4c}(\theta-2\pi n)^2\biggr]. \label{eqn:poissonsum}
\end{equation}
   To  prepare  the mock data, we add noise with  variance 
   $\delta \times P(Q)$ to the Gaussian  $P(Q)$. In  the analysis, we
   consider sets of  
   data  with various values of  $\delta$ and study the effects of 
$\delta$. \par
\begin{figure}
         \centerline{\includegraphics[width=10 cm, height=8
cm]{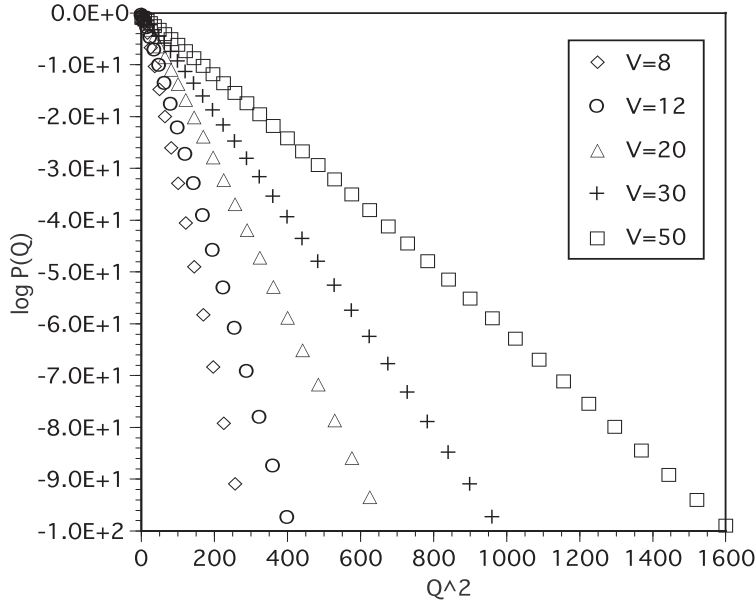}}
\caption{Gaussian topological charge distribution for  various
lattice volumes. The parameter $\delta$ is chosen to be 1/400. }
\label{fig:Pq}
\end{figure}
\begin{figure}
         \centerline{\includegraphics[width=10 cm, height=8
cm]{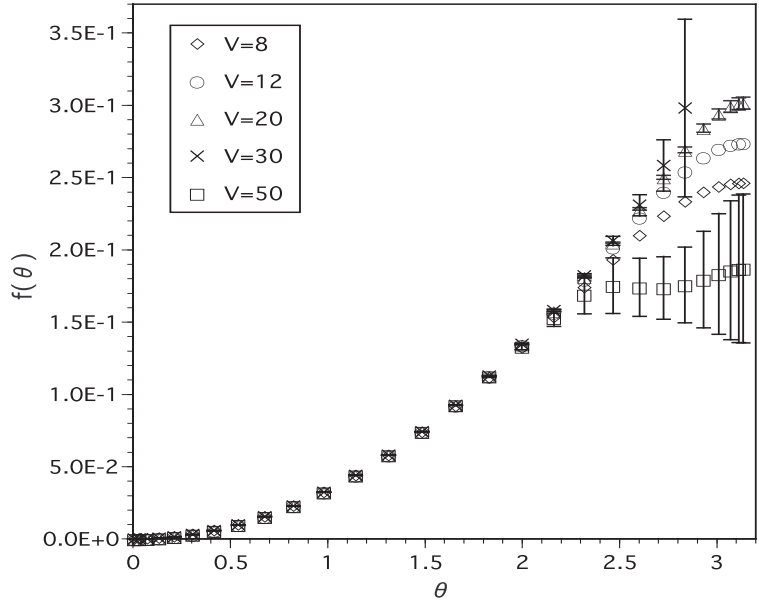}}
\caption{Free energy density $f(\theta)$ for  various lattice volumes.
$f(\theta)$ is calculated numerically by Fourier-transforming the
Gaussian topological charge distribution.
  }
\label{fig:free}
\end{figure}

   Figure~\ref{fig:Pq} displays the Gaussian $P(Q)$ for various lattice
   volumes $V$. The parameter $c$ is fixed at $c=7.42$. The corresponding
   free energy densities, $f(\theta)=-\frac{1}{V}\log{\cal Z}(\theta)$,
   calculated with  Eq.~(\ref{eqn:partitionfunction}), are plotted  in
   Fig.~\ref{fig:free}. All functions  $f(\theta)$ in Fig.~\ref{fig:free} fall on a
universal curve  for $\theta\siml 2.3$. For $\theta\simg 2.3$, finite size 
effects are
   clearly  observed.  As the  volume increases, $f(\theta)$ increases until
   $V\siml 20$,  but for $V=30$ and $V=50$,  the Fourier transformation does 
not work .
   At  $V=30$,  ${\cal Z}(\theta)$
   becomes negative for certain values of $\theta$,  and
   at $V=50$, $f(\theta)$ becomes almost flat for
   $\theta\simg 2.3$.
The latter behavior  causes  a fictitious signal of a first-order
   phase transition at $\theta\approx 2.3$.
   The mechanism of this  flattening~\cite{rf:PS,rf:IKY} \ 
    is briefly summarized as follows. \par
The distribution $P(Q)$ obtained from  MC simulations can be
   decomposed into two parts, a true value and an error:
\begin{equation}
   P(Q)={\tilde P}(Q)+\Delta P(Q), \label{eqn:Pqdata}
\end{equation}
   where $P(Q),\;{\tilde P}(Q)\;\mbox{and}\;\Delta P(Q)$ denote the MC data,
   the true value and the error, respectively.
In order to calculate $ P(Q)$ efficiently,  which ranges over many orders,
  the set method  and   the trial function method are conventionally used.
In the set method,  the range of $Q$ is divided into
several sets,  each of which consists of  several  bins.
In  each set, the topological charge of the
configurations is calculated in order   to construct  the histogram.
The trial function method makes the distribution of the histogram
almost flat. This is useful for reducing the error in   $ P(Q)$.
Accordingly, the error is  computed as~\cite{rf:HITY}
\begin{equation}
   \Delta P(Q)\sim \delta P(Q)\times P(Q), \label{eqn:Pqerror} \nonumber
\end{equation}
where $\delta P(Q)$ is   almost independent of $Q$.
Because  $P(Q)$ is a  rapidly decreasing function,  so is
   the statistical error $\Delta P(Q)$.
Thus,  the dominant contribution to  $\Delta P(Q)$ comes from that at
$Q=0$, and
   the partition function ${\cal Z}(\theta)$ has an  error
   $\Delta P(0)\approx \delta P(0)P(0)\approx \delta P(0)$.
   The free energy density for the MC data is
   approximately given by
\begin{eqnarray}
   f(\theta)&=&-\frac{1}{V}\log{\cal Z}(\theta) \approx
    -\frac{1}{V}\log\Bigl\{\sum_{Q}{\tilde P}(Q)e^{i\theta Q}+\delta 
P(0)\Bigr\}\nonumber\\
   &=&-\frac{1}{V}\log\bigl\{e^{-V{\tilde f}(\theta)}+\delta P(0)\bigr\},
    \label{eqn:freeenergydata}
\end{eqnarray}
where ${\tilde f}(\theta)$ is the true free energy density.
   The value of $e^{-V{\tilde f}(\theta)}$ decreases rapidly as the volume $V$
   and/or $\theta$ increase (see Fig.~\ref{fig:free}), and  the magnitude of
   $e^{-V{\tilde f}(\theta)}$ becomes of 
   order of $\delta P(0)$ at $\theta\simeq\theta_{\rm f}$. Therefore
   $f(\theta)$ cannot be calculated precisely, 
    but  $f(\theta)\simeq{\rm const.}$ for
   $\theta\simg\theta_{\rm f}$.
  This is the reason for the flattening behavior for  $V=50$ in the 
  region 
   $\theta\simg \theta_{\rm f}\simeq 2.3$, 
   as shown in Fig.~\ref{fig:free}.
  A drastic reduction of $\Delta P(Q)$  is necessary in order to 
  properly estimate
  $f(\theta)$ for $\theta\simg\theta_{\rm f}$. In the case of  a large
   volume, however, this is hopeless,   because an exponentially increasing
   amount of data  is   needed. Therefore,  we do
   need some other way to calculate $f(\theta)$. \par

%% file: sec3.tex
\section{MEM}
\label{sec:MEM}
\par
   In this section we briefly explain the concept of the  MEM
    and the necessary procedures for the analysis in order to make
   this paper self-contained and to define  the notation.
   \par

\subsection{MEM based on Bayes' theorem}
\label{subsec:Bayes}
\par
   In an experiment or a numerical simulation, data are always noisy,  and
   the number of sets of  data  is finite. It is in principle impossible to
   reconstruct the true image from such data sets. Hence,  it is reasonable
   to determine  the most probable image. From
    Bayes' theorem, the probability that an  image ${\bf f}$ occurs
   for  a given data set $\{{\bf D}\}$ is given  by
\begin{equation}
  {\rm prob}({\bf f}|{\bf D})=\frac{{\rm prob}({\bf D}|{\bf f}){\rm 
prob}({\bf f})}
    {{\rm prob}({\bf D})}, \label{eqn:Bayestheorem}
\end{equation}
   where ${\rm prob}(A)$ is the probability that an
   event $A$ occurs and ${\rm prob}(A|B)$ is the conditional probability
   that $A$ occurs under the  condition that $B$ occurs. Moreover, one can add to
   Eq.~(\ref{eqn:Bayestheorem}) `prior information' $I$ about the image
   ${\bf f}$.  The information $I$ includes that obtained from
   theoretical restrictions as well as knowledge based on  previous
   experiments. With $I$,  Eq.~(\ref{eqn:Bayestheorem}) becomes
\begin{equation}
  {\rm prob}({\bf f}|{\bf D},I)=\frac{{\rm prob}({\bf D}|{\bf f},I){\rm 
prob}({\bf f}|I)}
    {{\rm prob}({\bf D}|I)}, \label{eqn:Bayestheorem2}
\end{equation}
   where ${\rm prob}(A,B)$ is the joint probability that events $A$ and $B$
   occur simultaneously. The probability ${\rm prob}({\bf f}|{\bf D},I)$ is
   called the posterior probability. When the probability
   ${\rm prob}({\bf D}|{\bf f},I)$ is considered as a function of ${\bf f}$ for
   fixed data, it is equivalent to the likelihood function, which
   expresses how data points vary around the `true value' corresponding to
   the true image ${\bf f}$. The probability ${\rm prob}({\bf f}|I)$ is
   called the prior probability and represents our state of knowledge
   about the image ${\bf f}$ before the experiment is carried out. The most
   probable image satisfies the condition
\begin{equation}
   \frac{\delta~{\rm prob}({\bf f}|{\bf D},I)}{\delta\bf {f}}=0.
    \label{eqn:memcondition}
\end{equation}
   \par
   Recently, the MEM has been  applied to hadronic spectral functions in lattice
   QCD.~\cite{rf:AHN,rf:Yamazaki_CP-PACS} \ 
   In the analysis of the spectral function $A(\omega)$, the correlation
   function $D(\tau)$ is given by
\begin{equation}
   D(\tau)=\int_{0}^{\infty}d\omega K(\tau,\omega)A(\omega),
    \label{eqn:correlationfunction}
\end{equation}
   where $K(\tau,\omega)$ denotes the kernel of the Laplace
   transform. In lattice theories, the number of  data points
   is, at most, of  order 
   of $10 - 10^{2}$, due to the  finite volume, while the number of the
   degrees of freedom to describe the continuous function $A(\omega)$
   is in the 
   range ${\cal O}(10^2) - {\cal O}(10^3)$.
   \par
   Considering  theories that include the $\theta$ term,  what  we have to deal
   with is
\begin{equation}
   P(Q)=\int_{-\pi}^{\pi}d\theta \frac{e^{-i\theta Q}}{2\pi}{\cal Z}(\theta).
    \label{eqn:Pqint}
\end{equation}
   Comparing this with  Eq.~(\ref{eqn:correlationfunction}), we 
   see the correspondence
\begin{equation}
   \{P(Q)\leftrightarrow D(\tau),\;e^{-i\theta Q}/2\pi\leftrightarrow
    K(\tau,\omega),\;{\cal Z}(\theta)\leftrightarrow A(\omega)\};
    \label{eqn:correspondeces}
\end{equation}
   that is, the continuous function ${\cal Z}(\theta)$ must be reconstructed
   from a  finite number of data for $P(Q)$, which constitutes  an
   ill-posed problem. Given this situation, what we would like to do in the present
    paper is
    to rely on the  MEM,  employing the formula
\begin{equation}
   {\rm prob}({\cal Z}(\theta)|P(Q),I)={\rm prob}({\cal Z}(\theta)|I)
    \frac{{\rm prob}(P(Q)|{\cal Z}(\theta),I)}{{\rm prob}(P(Q)|I)}.
    \label{eqn:BayestheoremforPq}
\end{equation}
   \par
   The likelihood function ${\rm prob}(P(Q)|{\cal Z}(\theta),I)$ is given by
\begin{equation}
   {\rm prob}(P(Q)|{\cal Z}(\theta),I)=\frac{e^{-\frac{1}{2}\chi^{2}}}{X_L},
    \label{eqn:likelihoodfunction}
\end{equation}
   where $X_L$ is
   a normalization constant, and $\chi^{2}$  is defined 
   by
\begin{equation}
   \chi^2\equiv \sum_{Q,Q^\prime}(P^{({\cal Z})}(Q)-{\bar P}(Q))
    C^{-1}_{Q,Q^\prime}
    (P^{({\cal Z})}(Q^\prime)-{\bar P}(Q^\prime))
    \label{eqn:chisquare}
\end{equation}
   in our case, where $P^{({\cal Z})}(Q)$ is constructed from
   ${\cal Z}(\theta)$ through Eq.~(\ref{eqn:Pqint}). Here,  ${\bar
   P}(Q)$ denotes the average of a data set $\{P(Q)\}$, i.e., 
\begin{equation}
   {\bar P}(Q)=\frac{1}{N_d}\sum_{l=1}^{N_d}P^{(l)}(Q), \label{eqn:average}
\end{equation}
   where $N_d$ represents the number of sets of  data. The matrix
   $C^{-1}$ represents  the  inverse covariance obtained  from  the data set
   $\{P(Q)\}$.
   \par
   The prior probability ${\rm prob}({\cal Z}(\theta)|I)$ is given  in
   terms of the entropy $S$ as 
\begin{equation}
   {\rm prob}({\cal Z}(\theta)|I,\alpha)=\frac{e^{\alpha S}}{X_S(\alpha)},
    \label{eqn:priorprobability}
\end{equation}
   where $\alpha$ is a real positive parameter and $X_S(\alpha)$ denotes
   an $\alpha$-dependent normalization constant. The choice of the
   entropy $S$ is somewhat flexible. Conventionally, the
   Shannon-Jaynes entropy,
\begin{equation}
   S=\int^\pi_{-\pi} d\theta \biggl[{\cal Z}(\theta)-m(\theta)-
    {\cal Z}(\theta)\log\frac{{\cal Z}(\theta)}{m(\theta)}\biggr],
    \label{eqn:SJentropy}
\end{equation}
    is employed, 
   where $m(\theta)$ is called the `default model'. The default model
   $m(\theta)$ must be taken so as to be consistent with prior
   knowledge.
   Therefore the posterior probability
   ${\rm prob}({\cal Z}(\theta)|P(Q),I,\alpha,m)$ can be  rewritten as
\begin{equation}
   {\rm prob}({\cal Z}(\theta)|P(Q),I,\alpha,m)=\frac{e^{-\frac{1}{2}
    \chi^{2}+\alpha S}}{X_L X_S(\alpha)},
    \label{eqn:posteriorprobability}
\end{equation}
   where it is explicitly expressed  that $\alpha$ and $m$ are regarded as new
   prior knowledge in ${\rm prob}({\cal Z}(\theta)|P(Q),I,\alpha,m)$.
   \par
   The information $I$ restricts the regions to
   be searched in image space and helps us to effectively determine a 
   solution. We impose the  criterion
\begin{equation}
   {\cal Z}(\theta)>0,  \label{eqn:criterion}
\end{equation}
   so that ${\rm prob}({\cal Z}(\theta)\leq 0|I,m)=0$.
   \par
   In order to obtain the best image of ${\cal Z}(\theta)$, we must find
   the solution such that the function
\begin{equation}
   W\equiv -\frac{1}{2}\chi^{2}+\alpha S
    \label{eqn:functionw}
\end{equation}
   is maximized for a given $\alpha$:
\begin{equation}
   \frac{\delta}{\delta{\cal Z}(\theta)}(-\frac{1}{2}\chi^2+\alpha S)
    \Bigm|_{{\cal Z}={{\cal Z}^{(\alpha)}}}=0. \label{eqn:maximumcondition}
\end{equation}
   The parameter $\alpha$ plays the  role of determining the relative 
   weights of   $S$ and
   $\frac{1}{2}\chi^2$. For $\alpha=0$, the solution of
   Eq.~(\ref{eqn:maximumcondition}) corresponds to the maximal
   likelihood, while for $\alpha\gg 1$,  ${\cal Z}(\theta)=m(\theta)$ is
   realized as a solution. Therefore, care must be taken in  the choice
   of $m(\theta)$.
   \par

\subsection{Procedure for the analysis}
\label{subsec:procedure}
\par
   In the numerical analysis, the continuous function ${\cal Z}(\theta)$
   is discretized:\\
   ${\cal Z}(\theta)\to{\cal Z}(\theta_n)\equiv{\cal Z}_n$. Therefore, 
   the integral
   over $\theta$ in Eq.~(\ref{eqn:Pqint}) is converted into a  finite
   summation over $\theta$:
\begin{eqnarray}
   P_j=\left\{
       \begin{array}{ll}
        \displaystyle{\sum_{n=1}^{N_\theta}\frac{1}{2\pi}{\cal Z}_n}
	\;\;\;\;\;\;\;(j=1) \\
        \displaystyle{\sum_{n=1}^{N_\theta}\frac{\cos\theta_n j}{\pi}
	{\cal Z}_n}\;(\mbox{otherwise})
       \end{array}
      \right\}
   \equiv\sum_{n=1}^{N_\theta}K_{jn}{\cal Z}_n,  \label{eqn:Pqintdis}
\end{eqnarray}
   where $j=1,2,\cdots,N_q$. Note that $N_q< N_\theta$. Here, $P_j$ denotes
   $P(Q)$ at $Q=j-1$. Also,  we have used the fact that $P(Q)$ and ${\cal
   Z}(\theta)$ are even functions of $Q$ and $\theta$, respectively.
   Equation~(\ref{eqn:SJentropy}) is also discretized as
\begin{equation}
   S= \sum_{n=1}^{N_\theta}\biggl[{\cal Z}_{n}-m_{n}-
    {\cal Z}_{n}\log\frac{{\cal Z}_{n}}{m_{n}}\biggr],
     \label{eqn:SJentropydisorg}
\end{equation}
   where $m(\theta_n)\equiv m_n$.
   \par
   We employ  the following procedure for our analysis.~\cite{rf:Bryan,rf:AHN}
\vspace*{4mm}
\begin{enumerate}
   \item Maximizing $W$ for fixed $\alpha$: \par
\hspace*{5mm}
         In order to find the image that maximizes $W$ in the functional
         space of ${\cal Z}_n$ for a given $\alpha$, we calculate
         Eq.~(\ref{eqn:maximumcondition}). This yields
\begin{equation}
   -\alpha\log\frac{{\cal Z}_{n}}{m_{n}}=\sum_{i,j=1}^{N_q}K_{in}
    C^{-1}_{ij}\delta P_j, \;\;(n=1,2,\cdots,N_\theta)
    \label{eqn:memequationorg}
\end{equation}
	where $C_{ij}$ is the covariance matrix in
	Eq.~(\ref{eqn:chisquare}), and
\begin{equation}
  \delta P_j\equiv P_j^{({\cal Z})}-\bar{P}_j. \label{eqn:differenceofPq}
\end{equation}
\par
\hspace*{5mm}
	Solving for ${\cal Z}_n$ is non-trivial, because $N_\theta\sim
	{\cal O}(10^1) - {\cal O}(10^2)$ and $N_q\sim{\cal O}(10^0)$ in our
	case. It is convenient to use the SVD and  Newton's method. 
	Details are given  in Appendix~\ref{sec:apdx1}. The solution to
	 Eq.~(\ref{eqn:memequationorg}) is called ${\cal Z}^{(\alpha)}_n$. 
\vspace*{4mm}
\item  Averaging ${\cal Z}^{(\alpha)}_n$: \par
\hspace*{5mm}
         Since $\alpha$ is an artificial parameter, the final image 
         that we obtain 
         must have no $\alpha$ dependence. The $\alpha$-independent final image
         can be calculated by averaging the image ${\cal Z}^{(\alpha)}_n$
	with respect to  the probability. The expectation value of ${\cal Z}_n$
	is given by 
\begin{equation}
   {\hat {\cal Z}}_n=\int[d{\cal Z}]{\cal Z}_n\;
    {\rm prob}({\cal Z}_n|P(Q),I,m), \label{eqn:averageofZ}
\end{equation}
         where the measure
         $[d{\cal Z}]\equiv\prod_n d{\cal Z}_n/\sqrt{{\cal Z}_n}$ is
         used.~\cite{rf:Bryan} \ 
         Using the laws  of the total probability and the conditional
	probability, we obtain
\begin{eqnarray}
   {\hat {\cal Z}}_n&=&\int[d{\cal Z}]{\cal Z}_n\int d\alpha~
    {\rm prob}({\cal Z}_n|\alpha,P(Q),I,m){\rm prob}(\alpha|P(Q),I,m) \nonumber \\
   &\simeq&\int d\alpha~{\rm prob}(\alpha|P(Q),I,m){\cal Z}^{(\alpha)}_n.
\end{eqnarray}
      Further application  of the total probability, the conditional
	probability and  Bayes' theorem to ${\rm prob}(\alpha|P(Q),I,m)$ yields
	\begin{equation}
   {\hat {\cal Z}}_n=\frac{1}{X_W}\int d\alpha~{\cal Z}
   ^{(\alpha)}_n\exp\biggl\{\Lambda(\alpha)+W({\cal
     Z}^{(\alpha)})\biggr\},
     \label{eqn:averageofZ2}
\end{equation}
   where $X_{{\rm W}}$ is a normalization constant and
   $\Lambda(\alpha)\equiv
         \frac{1}{2}\sum_k\log\frac{\alpha}{\alpha+\lambda_k}$.
  	Here,    the values $\lambda_k$ are eigenvalues of the real symmetric
         matrix in $\theta$ space, 
         \begin{equation}
   \frac{1}{2}\sqrt{{\cal Z}_m}
    \frac{\partial^2\chi^2}{\partial{\cal Z}_m\partial{\cal Z}_n}
     \sqrt{{\cal Z}_n}\Bigm|_{{\cal Z}={\cal Z}^{(\alpha)}}.
     \label{eqn:matrixlambda}
\end{equation}
In deriving Eq.~(\ref{eqn:averageofZ2}),
       we have assumed that the probability
	${\rm prob}({\cal Z}_n|\alpha,P(Q),I,m)$
         has a sharp peak around ${\cal Z}^{(\alpha)}_n$, and
$W({\cal Z}^{(\alpha)})$ denotes the value of $W$ for which 
${\cal Z}_n={\cal Z}^{(\alpha)}_n$.
  The derivation of Eq.~(\ref{eqn:averageofZ2}) is given   in
   Appendix~\ref{sec:apdx2}.
   \par    
\hspace*{5mm}
     In averaging over $\alpha$, we determine a range of $\alpha$ so
         that
         ${\rm prob}(\alpha|P(Q),I,m)\geq  \frac{1}{10} \times  {\rm prob}({\hat 
\alpha}|P(Q),I,m)$
         holds, where ${\rm prob}(\alpha|P(Q),I,m)$ is maximized at
         $\alpha={\hat \alpha}$. The normalization constant is chosen 
         such that
\begin{equation}
   \int^{\alpha_{\rm max}}_{\alpha_{\rm min}}d\alpha~{\rm prob}(\alpha|P(Q),I,m)=1.
    \label{eqn:normalization}
\end{equation}
\vspace*{4mm}

\item Error estimation:\par

\hspace*{5mm}
         One of the advantages of the MEM is that it allows us  to estimate the 
error of
         constructed images.
	Because  the errors in  ${\cal Z}_n$ at  different points could be
         correlated, the error estimation should be performed over some
         range $\Theta$ in $\theta$ space. This  range $\Theta$ is
	determined systematically by analyzing the Hessian
         matrix in $\theta$ space, 
\begin{equation}
   H_{m,n}\equiv\frac{\partial^2 W}{\partial{\cal Z}_m \partial{\cal Z}_n}
    \Bigm|_{{\cal Z}={\cal Z}^{(\alpha)}}.
\end{equation}
\hspace*{5mm}
         The uncertainty of the final output image ${\hat {\cal Z}}_n$
         is calculated as~\cite{rf:JGU,rf:AHN}
\begin{equation}
   \langle (\delta {\hat {\cal Z}}_n)^2\rangle\equiv\int d\alpha
    \langle(\delta {\cal Z}^{(\alpha)}_n)^2\rangle {\rm prob}(\alpha|P(Q),I,m),
    \label{eqn:errorestimation}
\end{equation}
         where
\begin{eqnarray}
   \langle(\delta{\cal Z}_{m}^{(\alpha)})^2 \rangle &\equiv&
    \frac{\int[d{\cal Z}]\int_\Theta d\theta_n d\theta_{n^\prime}
     \delta{\cal Z}_n\delta{\cal Z}_{n^\prime} {\rm prob}({\cal Z}_m|P(Q),I,m,\alpha)}
      {\int[d{\cal Z}]\int_\Theta d\theta_n d\theta_{n^\prime}
       {\rm prob}({\cal Z}_m|P(Q),I,m,\alpha)} \nonumber \\
   &\simeq&- \frac{1}{\int_\Theta d\theta_n d\theta_{n^\prime}}
    \int_\Theta d\theta_n d\theta_{n^\prime} \Bigl(\frac{\partial^2 W}
     {\partial{\cal Z}_n\partial{\cal Z}_{n^\prime}}
      \Bigm|_{{\cal Z}={\cal Z}^{(\alpha)}}\Bigr)^{-1}.
      \label{eqn:errorestimationforalpha}
\end{eqnarray}
See Appendix~\ref{sec:apdx2} for  details.
\end{enumerate}
\vspace*{4mm}\par
   In the  procedure described in this section, the uniqueness of
   the final image is guaranteed for $\alpha\neq 0$. This  requires the
   conditions that the image ${\cal Z}^{(\alpha)}(\theta)$ be 
   positive definite and  the kernel $(K^t)_{nj}$ be real. These are 
   indeed satisfied, as shown in Appendix~\ref{sec:apdx3} .
\par

%% file: sec4.tex
\section{Results}
\label{sec:result}\par
In this section, we present the
results  of the MEM analysis of the data for $P(Q)$.
	 To prepare    data for the analysis, we added to  $P(Q)$ Gaussian
noise generated with the variance
  $\delta\times P(Q)$ for each value of $Q$. This way of adding noise is
  based on the procedure which was employed to calculate
$P(Q)$ in the simulations of the CP$^{N-1}$
model.~\cite{rf:HITY}
This yields error which amounts to almost constant portion of
$P(Q)$ for each $Q$, as mentioned above Eq.~(\ref{eqn:Pqerror}).
The parameter  $\delta$ was varied  from 1/10 to 1/600, and we present
the results for $\delta=1/400$.  A set of data consists of $P(Q)$ along 
with the errors from  $Q=0$ to $Q=N_{q}-1$.
Employing $N_d$ such sets of
data, we calculated the covariance matrices in
Eq.~(\ref{eqn:chisquare}) with the jackknife method.  We have
checked whether the outcome is  stable by  varying   the
value of $N_d$ in the range $10\leq N_d  \leq 60$ and found
that this is the case for $30\siml  N_d $.  We present here the
results for      $N_d=30$. \par
  For the default model $m(\theta)$ in Eq.~(\ref{eqn:SJentropy}), we
studied   various  cases:
       (i) $m(\theta)=$ const., (ii)
$m(\theta)=(\sin(\theta/2)/(\theta/2))^{V}$ $\equiv m_{\rm
strg}(\theta)$ ,  (iii) $m(\theta)=\exp(-\frac{\ln 10}{\pi^2}  \gamma
\theta^2) $.  In case (i), we studied several values, 
$m(\theta)=0.1, 0.3$  and 1.0.
We present the results for $m(\theta)=1.0$ as a typical case.  Case
(ii) corresponds to the
       strong coupling limit of the ${\rm CP}^{N-1}$ model. Case
(iii) is the Gaussian case. The parameter $\gamma$ was  varied in the
analysis.\par
\begin{figure}
        \centerline{\includegraphics[width=12 cm, height=8
cm]{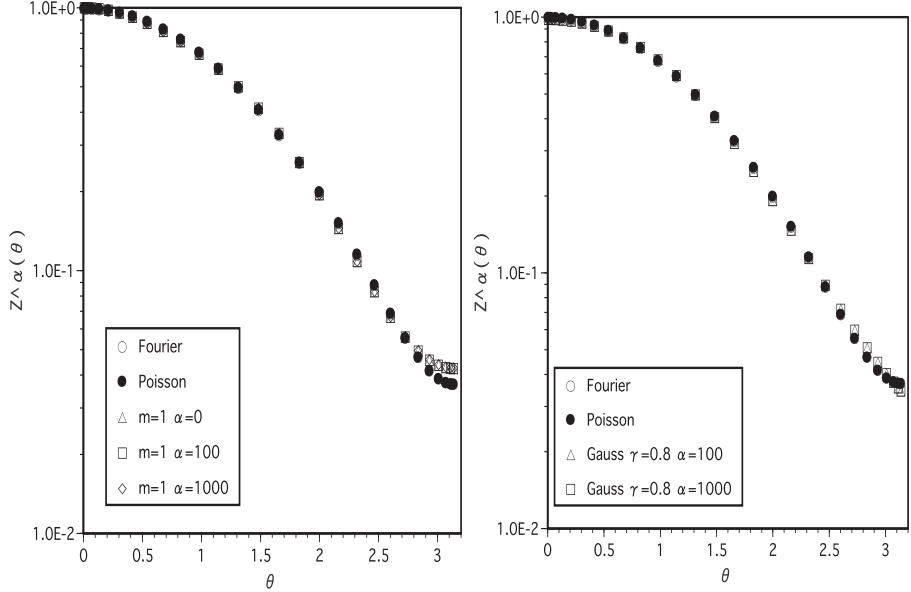}}
\caption{${\cal Z}^{(\alpha)}(\theta)$ for the data without flattening.
 Here, $V=12$.
The default
model $m(\theta)$ is chosen to be the constant 1.0 and the Gaussian
 function with $\gamma=0.8$.
}
\label{fig:MaxZ_nf}
\end{figure}
The number of degrees of freedom in $\theta$ space,  $N_\theta$, is
larger than that of the topological charge,  $N_q$.
The number $N_q$ was chosen so as to satisfy  $P(Q)\geq 10^{-30}$ in the
non-flattening case and $P(Q)\geq 10^{-11}$ in the flattening case,
and  it varies from 3 to 8, depending on $V$.
  The covariance matrix appears in Eq.~(\ref{eqn:chisquare}):
\begin{equation}
   C_{Q,Q^\prime}=\frac{1}{N_d ( N_d-1)}\sum_{l=1} ^{N_d}(
P^{(l)}(Q)-{\bar P}(Q))( P^{(l)}(Q^\prime)-{\bar P}(Q^\prime) ),
   \label{eqn:covariance}
\end{equation}
  where $P^{(l)}(Q)$ denotes the $l$-th data of the topological charge
distribution and ${\bar P}(Q)$ is the average Eq.~(\ref{eqn:average}).
The inverse covariance matrix is calculated with such precision
   that the  product of  the   covariance matrix  and its inverse has
off-diagonal elements that are at most ${\cal O}(10^{-27})$.
\par
  The number of the other degrees of freedom, $N_\theta$, was varied
  from 10 to 100, 
and it was found that the results are stable
  for $25 \siml  N_\theta$.  In the following results, $N_\theta$ is
set to be 28.
Note that in order to reproduce ${\cal Z}(\theta)$, which ranges over many
orders,  the analysis  must be  performed with  quadruple  precision.
\par

\begin{figure}
        \centerline{\includegraphics[width=9 cm, height=8
cm]{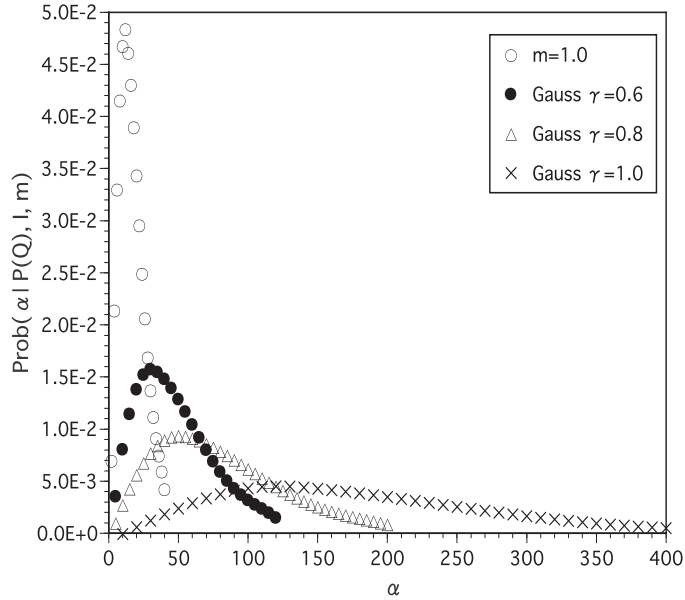}}
\caption{${\rm prob}(\alpha|P(Q),I,m)$ for the data without
 flattening. Here, $V=12$.
The default model is chosen to be the constant $m(\theta)=1.0$ and the
 Gaussian function with $\gamma=0.6$, 0.8 and 1.0.}
\label{fig:P_alp_nf}
\end{figure}
\begin{figure}
        \centerline{\includegraphics[width=10 cm, height=8
cm]{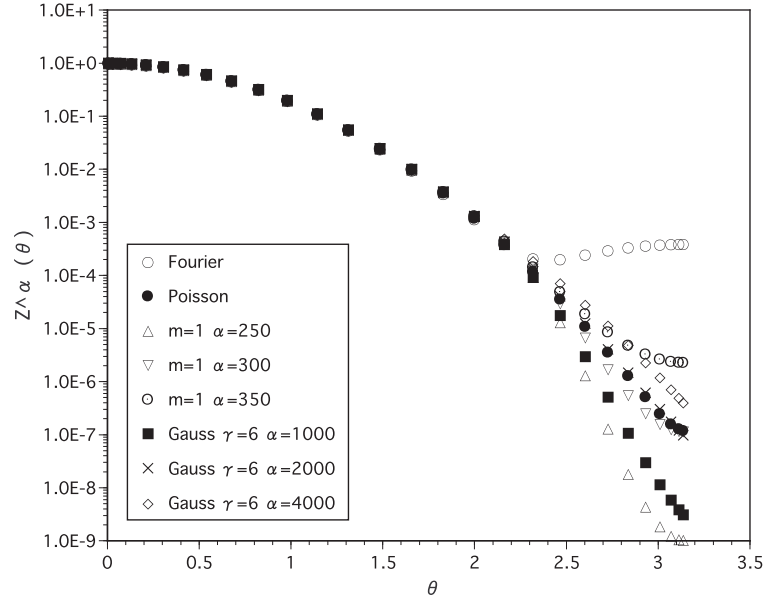}}
\caption{${\cal Z}^{(\alpha)}(\theta)$ for the data with flattening. The
default
model $m(\theta)$ is chosen to be the constant 1.0 and the Gaussian
 function with $\gamma=6$. 
Comparing with the result of the
Fourier transform (circles), the result of the MEM for certain values of
$\alpha$ ($\alpha\approx 300$ for $m(\theta)=1.0$ and   $\alpha\approx 2000$
for the Gaussian) approximately reproduces that of the exact partition function
Eq.~(\ref{eqn:poissonsum}) (filled circles). }
\label{fig:MaxZ_f}
\end{figure}
\subsection{MEM analysis of the data without flattening}
\label{sec:result_noflat}\par
Before discussing  to what extent  the flattening behavior of the
  free energy  is remedied, we discuss the case without
  flattening. It is a non-trivial question  whether the MEM is effective
  in the application considered here.
  \par
  The data for $V\leq 20$ were used in the analysis. For such data, no
  flattening behavior is observed (see Fig.~\ref{fig:free}). For the
  present, we
  concentrate on the data for $V=12$. For this value of $V$, two types
  of  default models were employed, the constant default model,
  $m(\theta)=1.0$, and the Gaussian one with $\gamma=0.4-1.0$.
  \par
  The maximal image ${\cal Z}^{(\alpha)}(\theta)$ of ${\cal Z}(\theta)$
  for a given $\alpha$ was calculated using Eq.~(\ref{eqn:maximumcondition}).
    Figure \ref{fig:MaxZ_nf} displays ${\cal Z}^{(\alpha)}(\theta)$
  calculated in this way 
  for various $\alpha$ ($0\leq \alpha \leq 1000$) in the cases that
  $m(\theta)$ is the constant 1.0 and the Gaussian function with
  $\gamma=0.8$. It is found
  that there is almost no discernible $\alpha$ dependence of ${\cal
  Z}^{(\alpha)}(\theta)$ and that the images approximately agree with
  the result of the Fourier 
  transform, and thus with the exact partition  function
  ${\cal Z}_{{\rm pois}}(\theta)$.
  \par
  In order to determine which ${\cal Z}^{(\alpha)}(\theta)$ is the
  most probable, we calculated  the posterior probability ${\rm
  prob}(\alpha|P(Q),I,m)$ 
   in Eq.~(\ref{eqn:averageofZ2}). The data for ${\rm
  prob}(\alpha|P(Q),I,m)$ were fitted 
  by smooth functions and normalized such that
  Eq.~(\ref{eqn:normalization}) holds.  The results are plotted in
  Fig.~\ref{fig:P_alp_nf}.  For both cases,
  ${\rm prob}(\alpha|P(Q),I,m)$ has a peak at a small value of $\alpha$;
  for $m(\theta)=1.0$, the peak is located at $\alpha\simeq 12.0$,
  while in the Gaussian case it is located at $\alpha\simeq 30.0$
  $(\gamma=0.6)$, $\alpha\simeq  
  50.0$ $(\gamma=0.8)$ and $\alpha\simeq 120.0 $ $(\gamma=1.0)$. Because
  the functions ${\cal Z}^{(\alpha)}(\theta)$ do not depend on 
  $\alpha$ in the region around the peak, the
  integrals we must evaluate to obtain the averaged image ${\hat{\cal
  Z}}(\theta)$ in 
  Eq.~(\ref{eqn:averageofZ2}) are trivially simple, and
  the functions ${\hat{\cal Z}}(\theta)$ are approximately in agreement
  with the exact one. 
  \par
  For $V=8$ and 20, similar analyses were carried out.
  We find that the  characteristics of ${\cal Z}^{(\alpha)}(\theta)$
  and ${\rm prob}(\alpha|P(Q),I,m)$ stated above, namely, that 
  ${\cal Z}^{(\alpha)}(\theta)$ is almost independent of $\alpha$ for
  not so large values ($\alpha\siml 1000$) and ${\rm
  prob}(\alpha|P(Q),I,m)$  has a
  peak  at a small value of $\alpha$, are also observed for  $V=8$ and 20.
  Therefore,  we obtain the same results for ${\hat{\cal Z}}(\theta)$.
  More generally,  in the non-flattening case,
  where the Fourier transform works well,
     the fact that the image obtained using the MEM is consistent with the
  result of  the Fourier transform can be understood by
 carefully considering the equations used in the SVD.
   This is investigated analytically  in  Appendix~\ref{sec:apdx4}.
   \par
   The detailed procedure for estimating the error $\delta {\hat{\cal 
Z}}(\theta)$
   is discussed in the next subsection, and its results are given at the
   end.
\par
\begin{figure}
         \centerline{\includegraphics[width=10 cm, height=8
cm]{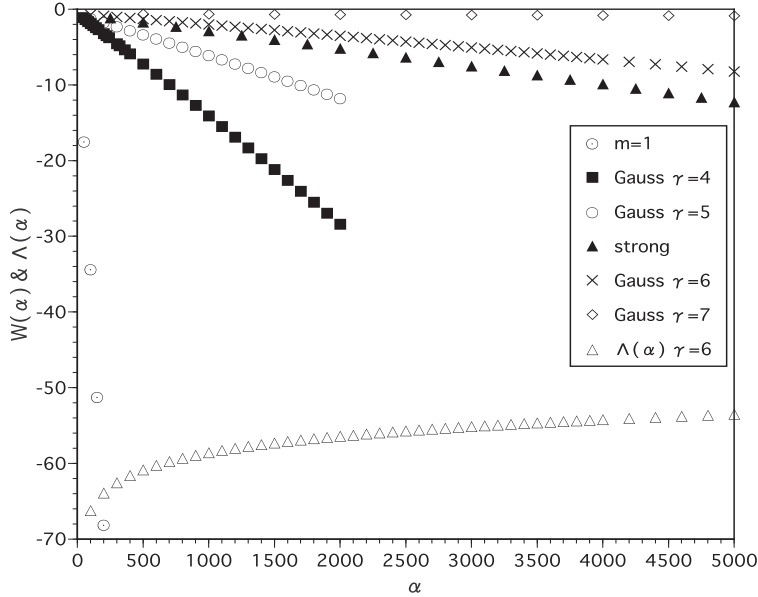}}
\caption{$W({\cal Z}^{(\alpha)})$ for various $m(\theta)$ and for  $V=50$.
  $W({\cal Z}^{(\alpha)})$ is plotted for $m_{\rm strg}(\theta)$ and for
 the Gaussian default models with $\gamma=4,5,6$ and 7.  $\Lambda(\alpha)$
 appearing in Eq.~(\ref{eqn:averageofZ2}) is also plotted on  the same scale.
   Because $\Lambda(\alpha)$ does not depend strongly on $m(\theta)$,  
 $\Lambda(\alpha)$ for the Gaussian default model with $\gamma=6$
is displayed as a reference. }
\label{fig:W_max}
\end{figure}
\subsection{MEM analysis of the data with flattening}
Let us now turn to the case with flattening.
Unlike the case without  flattening, in this case the images of ${\cal
Z}(\theta)$ display behavior that differs greatly from those   of the
Fourier transform. 
We fix the volume to $V=50$  for the time being.
Figure \ref{fig:MaxZ_f} displays  ${\cal Z}^{(\alpha)}(\theta)$ calculated
with   $m(\theta)$ given by the constant 1.0 and the Gaussian form with
$\gamma=6$, 
which are  images determined by maximizing Eq.~(\ref{eqn:functionw})
for each  $\alpha$.
For each  of the defaults,  the maximal images are free from  
flattening, and  at least for a certain value of $\alpha$, 
there exists a ${\cal Z}^{(\alpha)}(\theta)$ which 
is in reasonable agreement with the exact one, ${\cal Z}_{\rm pois}(\theta)$.
  This also
holds in the case that $m(\theta)$ is Gaussian with  $\gamma=4,5$ and 7,
although the value of $\alpha$ for which we find the best agreement with
${\cal Z}_{\rm pois}(\theta)$ depends on $\gamma$.
In the case of $m_{\rm strg}(\theta)$, however,  we find  no
agreement, even when $\alpha$ is
varied from ${\cal O}$(1) to ${\cal O}(10^{6})$.   \par
\begin{figure}
        \centerline{\includegraphics[width=10 cm, height=8
cm]{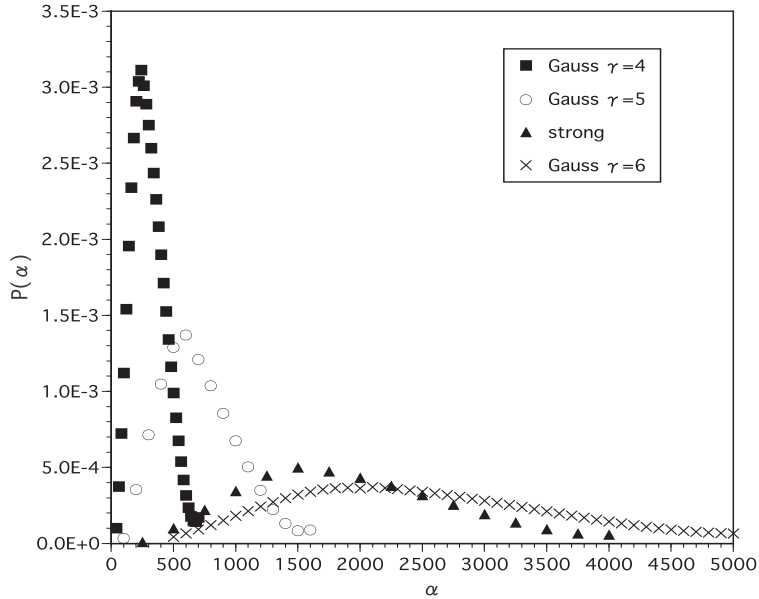}}
\caption{${\rm prob}(\alpha|P(Q), I,m)$ for various $m(\theta)$. $V$ is fixed
to 50.}
\label{fig:P_alp}
\end{figure}

The posterior probability ${\rm prob}(\alpha|P(Q), I,m)$ was calculated
with Eq.~(\ref{eqn:Palpha}) appearing in Eq.~(\ref{eqn:averageofZ2}).
Figure \ref{fig:W_max} displays the behavior of $W({\cal Z}^{(\alpha)})$
and $\Lambda(\alpha)$ for  various $m(\theta)$.
We find that as $\alpha$ increases, $W({\cal Z}^{(\alpha)})$ decreases
almost linearly,  depending  strongly  on
$m(\theta)$, while $\Lambda(\alpha)$ increases with rather weak 
$\alpha$ dependence. The sum of $W({\cal Z}^{(\alpha)})$  and
$\Lambda(\alpha)$ gives  ${\rm prob}(\alpha|P(Q), I,m)$,  and the balance
between the two determines the location of the peak of
${\rm prob}(\alpha|P(Q),I,m)$, if it exists.
  This is shown in Fig.~\ref{fig:P_alp}.
\begin{figure}
        \centerline{\includegraphics[width=10 cm, height=8
cm]{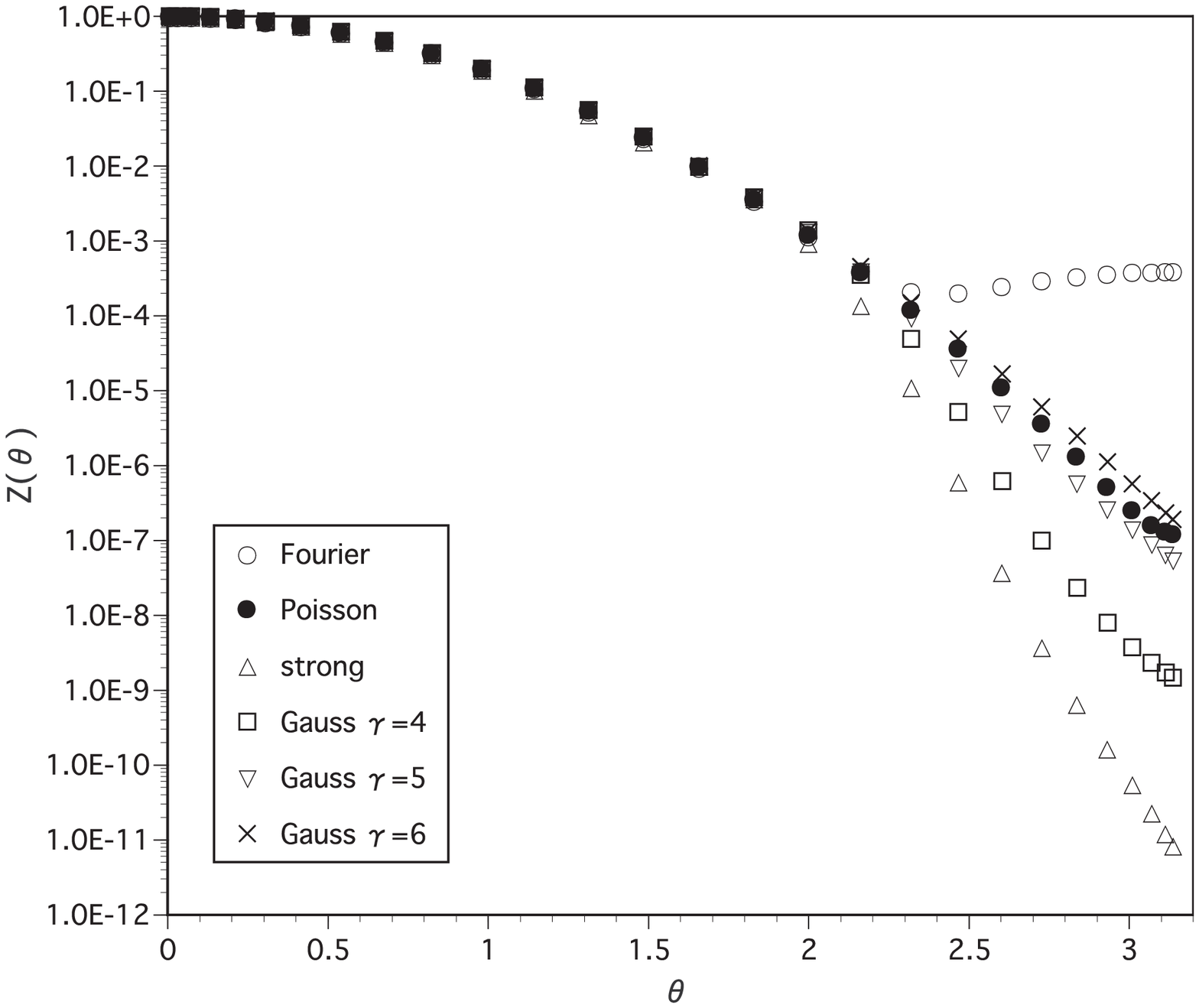}}
\caption{Averaged partition function  $\hat {{\cal Z}}(\theta)$ for various
$m(\theta)$. The volume is fixed to  50. The result of the Fourier
 transform is
also included. Those  for  the   Gaussian default models with  $\gamma=5$
and 6  agree reasonably with the exact
result, ${\cal Z}_{\rm pois}(\theta)$. The result for $m_{\rm strg}(\theta)$
shows a large deviation from ${\cal Z}_{\rm
pois}(\theta)$.  }
\label{fig:Z_av}
\end{figure}
\par
The averaged image $\hat {{\cal Z}}(\theta)$ was calculated using
Eq.~(\ref{eqn:averageofZ2}). Integrating over $\alpha$,
  the  data for ${\rm prob}(\alpha|P(Q), I,m)$ were fitted by a smooth
  function. Then, the fitting function was normalized.
Figure \ref{fig:Z_av} displays  $\hat {{\cal Z}}(\theta)$ for various
$m(\theta)$.
In the case of a Gaussian $m(\theta)$, the parameter $\gamma$ is
   varied from 4 to 8.  We find  reasonably good agreement
 between the function ${\hat{\cal Z}}(\theta)$ for $\gamma=5$ and 6.
 This is due to the fact that for these cases, 
the best value
of $\alpha$ for ${\cal Z}^{(\alpha)}(\theta)$  is approximately equal to the
location of the peak of
${\rm prob}(\alpha|P(Q), I,m)$;  for $\gamma=6$ the two values are
nearly equal($\approx$  2000), while  for $\gamma=5$
they  differ slightly ($\approx$ 1000 for the former and $\approx$ 
650 for the latter).
In other words, these images could occur  with   high probability.
\begin{figure}
        \centerline{\includegraphics[width=8 cm, height=6
cm]{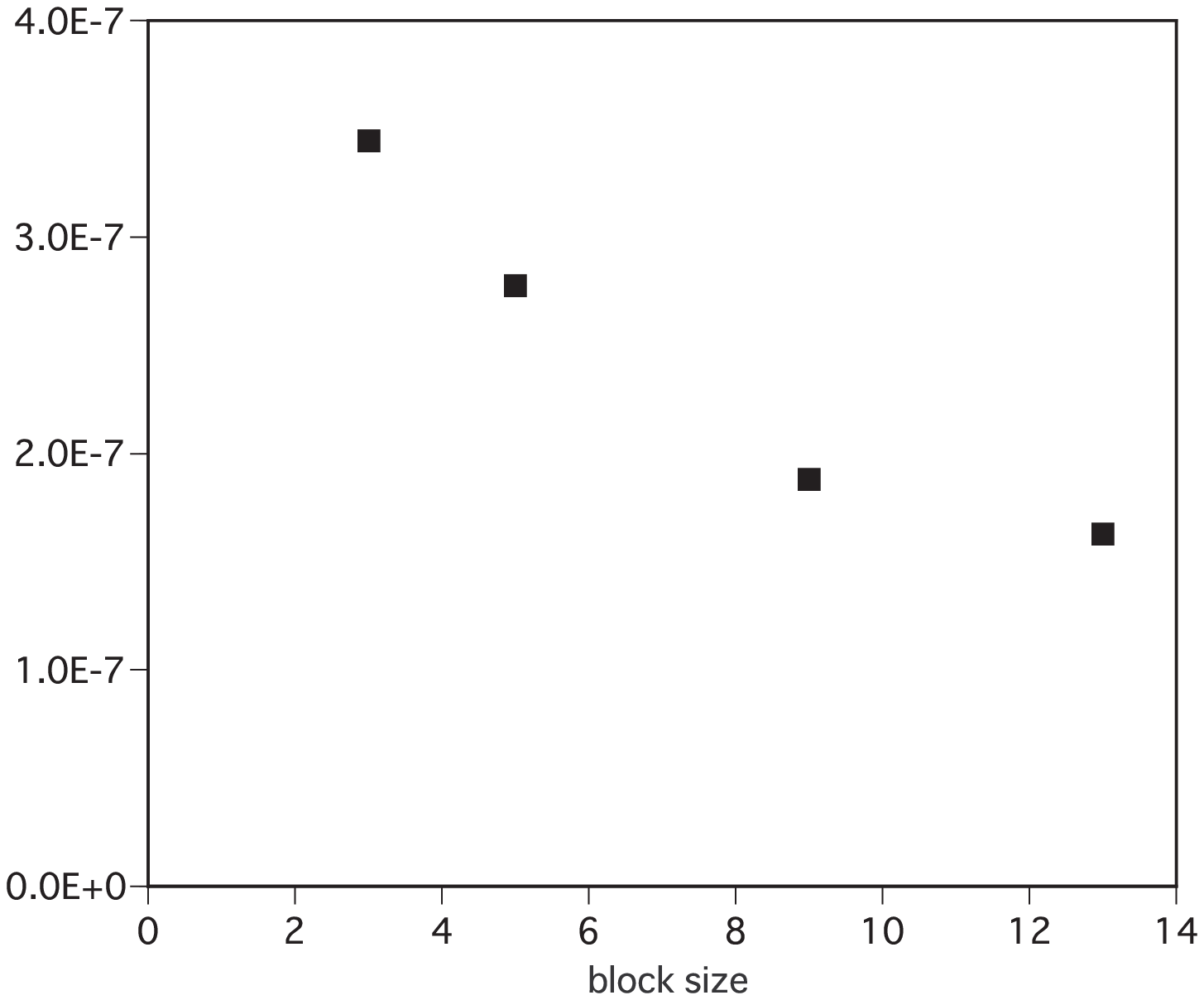}}
\caption{Errors estimated at $\theta=3.07$ by varying the block size
for $\gamma=5.5$ and $V=50$.
  (See the text for details.)}
\label{fig:error}
\end{figure}
\begin{table}
\caption{$\hat {{\cal Z}}(\theta)$  and error at $\theta=3.07$ for the various 
 default models depicted in Fig.~\ref{fig:Z_av}.  The exact value,   
${\cal  Z}_{\rm pois}(3.07)$, is $1.554\times 10^{-7}$.}
\label{table:m&Z&deltaZ}
\begin{center}
\begin{tabular}{cr@{.}lr@{.}l}
\hline
\hline
$m(\theta)$ &  
\multicolumn{2}{c}{  $\hat {{\cal Z}}(3.07)$ } &
   \multicolumn{2}{c}{$\delta \hat {{\cal Z}}(3.07)$}\\
\hline
strong  & 2&33$\times 10^{-11}$ & 3&5$\times 10^{-8}$  \\
Gauss $\gamma=4$  & 2&29$\times 10^{-9}$ & 7&4$\times 10^{-8}$ \\
Gauss $\gamma=5$  & 8&99$\times 10^{-8}$ & 1&9$\times 10^{-7}$\\
Gauss $\gamma=6$  & 3&38$\times 10^{-7}$ & 1&7$\times 10^{-7}$\\
\end{tabular}
\end{center}
\end{table}
In the $m(\theta)=1.0$ case, although the best image
is in good agreement with ${\cal Z}_{\rm pois}(\theta)$ for $\alpha=300$, as
shown in
Fig.~\ref{fig:MaxZ_f},  the peak of ${\rm prob}(\alpha|P(Q), I)$ is
located in the region where 
$\alpha < 40.0 $. This large difference  in $\alpha$ leads to    a
large deviation
of $\hat {{\cal Z}}(\theta)$ from the correct one,  ${\cal Z}_{\rm
pois}(\theta)$. Similarly, for $m_{\rm
strg}(\theta)$,  the image ${\cal Z}^{(\alpha)}(\theta)$
in agreement  with ${\cal Z}_{\rm
pois}(\theta)$ occurs with a very low probability,  and consequently
 $\hat {{\cal Z}}(\theta)$ deviates
greatly from ${\cal Z}_{\rm pois}(\theta)$.   This is obvious,   because
${\cal Z}^{(\alpha)}(\theta)$
never agrees with ${\cal Z}_{\rm pois}(\theta)$ for
  $\alpha={\cal O}(1)$ to ${\cal O}(10^{6})$.
(Note that the errors are not included in the figure.) 
 \par

One of the advantages of the MEM analysis is that it allows estimation
of  the
error according to the probability that the images are realized. In fact,
the obtained
images are meaningless unless their errors are evaluated.
We use the formula Eq.~(\ref{eqn:errorestimationforalpha}) for the
error
  estimates. The error $\delta \hat {{\cal Z}}(\theta_n)$
  of $\hat {{\cal Z}}(\theta_n)$ is estimated  by integrating $\theta$ in
  Eq.~(\ref{eqn:errorestimationforalpha})
  over the range $\Theta=\theta_{n-b/2}$ and $\theta_{n+b/2}$, where
  $\theta_m$ is the abscissa in the Gauss-Legendre $N$-point ($N$=100)
  quadrature formula for the range $0\leq\theta\leq\pi$.
 Figure \ref{fig:error} displays typical behavior of the error as a
function of the block size at $\theta=3.07$ for Gaussian $m(\theta)$
 with $\gamma=5.5$, where the block size is defined by $b+1$.  
\begin{figure}
        \centerline{\includegraphics[width=10 cm, height=8
cm]{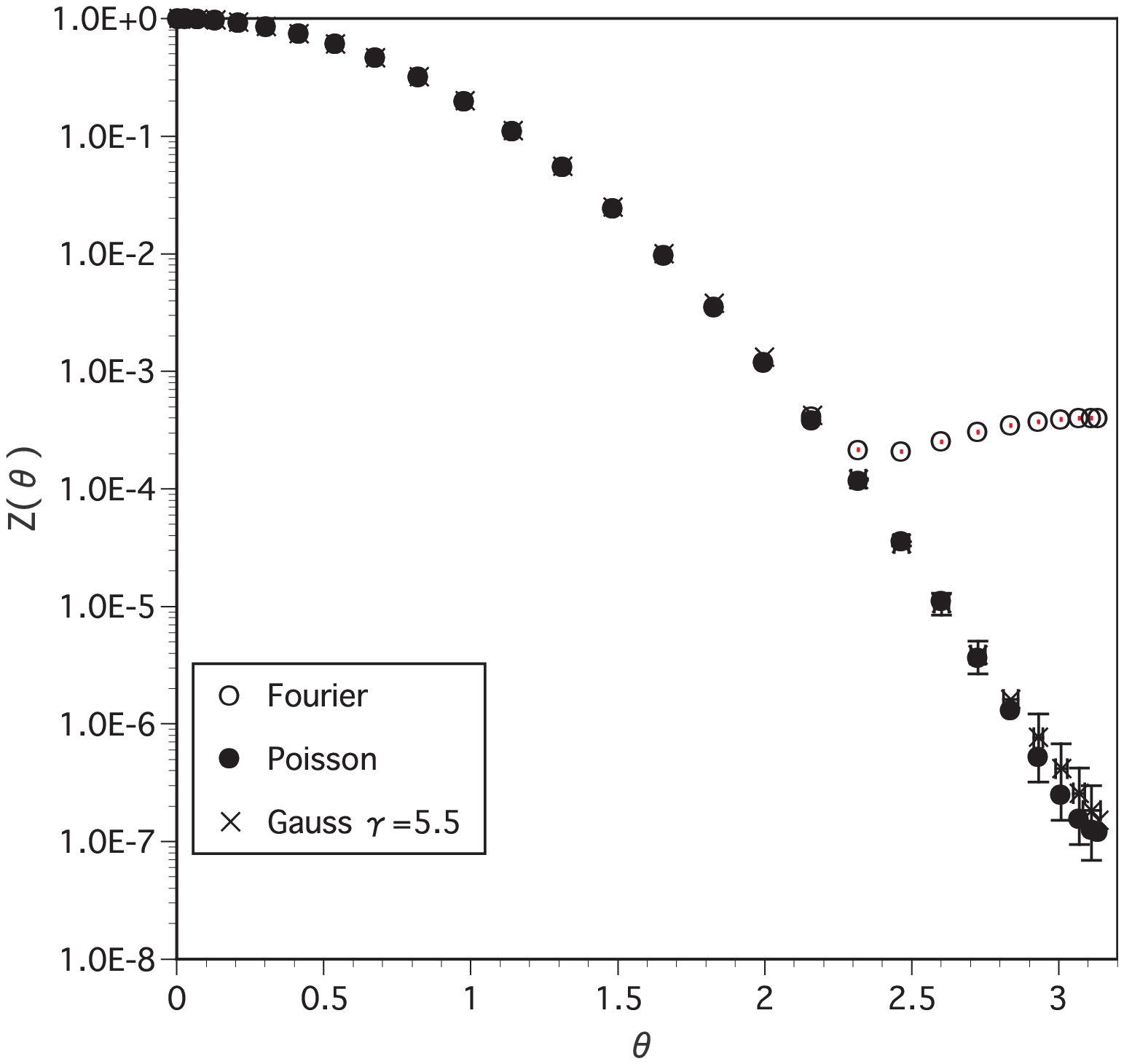}}
\caption{$\hat {{\cal Z}}(\theta)$ (crosses) with the error bars  for  the
Gaussian default model with $\gamma=5.5$. Here, $V=50$. Compared to the
 result of
 the Fourier transform (circles), 
a remarkable improvement is clearly seen.  }
\label{fig:Z_quad}
\end{figure}
\begin{figure}
        \centerline{\includegraphics[width=10 cm, height=8
cm]{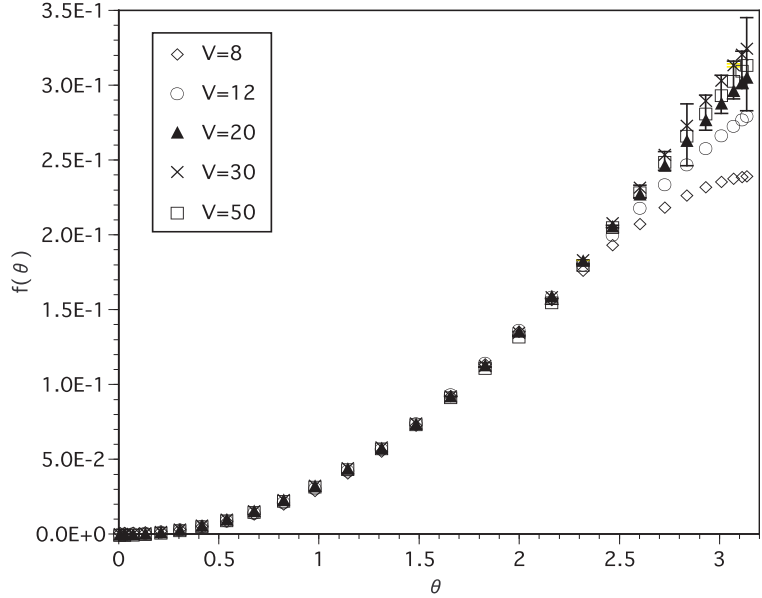}}
\caption{Free energy density $f(\theta)$ calculated from
$\hat {{\cal Z}}(\theta)$
for various
volumes.  (See Fig.~\ref{fig:free} for a
comparison.)}
\label{fig:free_MEM}
\end{figure}
\begin{table}
\caption{$\hat {{\cal Z}}(\theta)$  at $\theta=2.30$ and
$\theta=3.07$ for  various volumes $V$. The default model  is chosen to be the
Gaussian form with $\gamma$. The exact values ${\cal  Z}_{\rm pois}(\theta)$
are also listed.}
\label{table:V&Z&deltaZ}
\begin{center}
\begin{tabular}{ccr@{.}lr@{.}lr@{.}lr@{.}lr@{.}l}
\hline
\hline
$V$ & $\gamma$ (Gaussian default) & \multicolumn{2}{c}{ ${\cal  Z}_{\rm
pois}(2.30)$} &
\multicolumn{2}{c}{ $\hat {{\cal Z}}(2.30)$ } &
\multicolumn{2}{c}{  ${\cal  Z}_{\rm pois}(3.07)$ } &
   \multicolumn{2}{c}{$\hat {{\cal Z}}(3.07)$}\\
\hline
8 &   0.1 & 2&493$\times 10^{-1}$ & 2&424(3)$\times 10^{-1}$ &
   1&407$\times 10^{-1}$ & 1&48491(6)$\times 10^{-1}$ \\
12  &  0.8 & 1&155$\times 10^{-2}$ & 1&145(3)$\times 10^{-2}$ &
   3&752$\times 10^{-2}$ & 3&762(1)~~~$\times 10^{-2}$ \\
20 &   1.6  & 2&676$\times 10^{-2}$ & 2&583(2)$\times 10^{-2}$ &
   2&697$\times 10^{-3}$ & 2&946(6)~~~$\times 10^{-3}$ \\
30 & 3.4 & 4&372$\times 10^{-3}$ & 4&12(2)$\;\;\times 10^{-3}$ &
   1&023$\times 10^{-4}$ & 8&3(3)~~~~~~$\times 10^{-4}$ \\
50 & 5.5  & 1&169$\times 10^{-4}$ & 1&20(4)$\;\;\times 10^{-4}$ &
   1&554$\times 10^{-7}$ & 2&5(1.6)~~$\;\;\times 10^{-7}$ \\
\end{tabular}
\end{center}
\end{table}
In order to show how large the errors of $\hat {{\cal Z}}(\theta)$  in Fig.~\ref{fig:Z_av} 
are in the region near $\theta=\pi$,   Table~\ref{table:m&Z&deltaZ}
lists the values of 
$\hat {{\cal Z}}(\theta)$ and their errors at $\theta=3.07$.
 Together with the results displayed in 
Fig.~\ref{fig:Z_av}, we find  that  
 the more $\hat {{\cal Z}}(\theta)$ deviates from the  exact one,
 the larger the magnitude of the relative error becomes. 
\par
Among the default models we have investigated,  it turns out that the 
errors for the Gaussian form with $\gamma=5.5$ are the smallest. Figure
\ref{fig:Z_quad} plots the
partition function  with the  estimated 
errors for the Gaussian default model with $\gamma=5.5$. The result is
consistent with ${\cal Z}_{\rm pois}(\theta)$ and does not
exhibit flattening. For the other default models, $m(\theta)=1.0$ and
$m_{\rm strg}(\theta)$, it is found
that the errors are too large to obtain reasonable images ${\hat{\cal
Z}}(\theta)$. 
 \par
We applied  the same procedure to  the data for other volumes.
Table~\ref{table:V&Z&deltaZ} gives  $\hat {{\cal Z}}(\theta)$
  and  $\delta \hat {{\cal Z}}(\theta)$ at $\theta=2.30$ and 3.07 for
  various volumes and default models. Each of the two values of $\theta$
  was 
  chosen as a reference, where the latter is close to $\pi$ and the
  former is near $\theta_{\rm f}$. 
  The default models listed in Table~\ref{table:V&Z&deltaZ} give almost
  minimal errors among  
  the defaults  we have investigated for each volume. These were used 
  to calculate the free  energy density plotted in 
  Fig.~\ref{fig:free_MEM}.  
 We find by comparing Fig.~\ref{fig:free} that the flattening behavior
 is no longer observed.

%% file: sec5.tex
\section{Summary}
\label{sec:summary}\par
    We have considered  lattice field theory with a $\theta$ term. When
    studied numerically, this theory suffers from
the complex Boltzmann weight problem, or the sign problem.
  As an attempt at  an approach that differs from  the conventional
procedure, which employs the numerical Fourier transform of  $P(Q)$, we have
applied   the  MEM in order  to reconstruct  the partition function
${\cal Z}(\theta)$.  
In the MEM analysis, the image of ${\cal Z}(\theta)$ is calculated 
probabilistically, and the error is estimated as the uncertainty 
of the image.  
We   have   employed  the Gaussian $P(Q)$ as a test, 
because its
Fourier transform can be computed analytically. We found that for the
  data without flattening,
the results of the MEM analysis are consistent with those of the
Fourier transform, while for those with flattening,   the MEM
 reproduces reasonable images that exhibit no flattening
 by reducing the error contained in the data of $P(Q)$ ($\delta\approx 1/400$). 
\par
  Comments are in order.
\begin{enumerate}
\item During the analysis,
   the condition ${\cal Z}(\theta) >0$ in  Eq.~(\ref{eqn:criterion})
was imposed as prior  information  $I$.
This  plays an important role in searching for the maximal image
  in the case with flattening.
  This condition could yield results that differ from those found in the
      Fourier method, where
${\cal Z}(\theta)$ could become negative due to large error.
\item It is  important what default model is  chosen  for the
analysis.
A criterion for this choice is the magnitude of the  errors of the
averaged image, which are determined  according to the
probability. We have investigated various default models
  and found that  $\gamma =3.4$ for $V=30$ and
  $\gamma =5.5$ for $V=50$ are  the best among those which  we have
  investigated in the flattening case. The purpose of the present paper
      is to check the feasibility of the application of the MEM to the
      case with the $\theta$ term, and therefore we did not consider a
      large number of Gaussian defaults with different values of
      $\gamma$. 
There might be some  better default models than those we have
investigated here.
\item  The final image ${\cal Z}(\theta)$ depends on $\delta$, 
which controls   the magnitude of  the error of  the mock data 
$P(Q)$. The  parameter $\delta$ directly affects  the covariance 
matrix and indirectly influences  the error $\delta \hat {{\cal Z}}(\theta)$  
of the image, which  
is  the uncertainty of the image.   Although we employed the least uncertainty as a 
criterion, it is still unclear to what extent the obtained best 
image is close to the true one; i.e., there is  a systematic 
error in $\hat{{\cal Z}}(\theta)$, not reflected in 
$\delta \hat {{\cal Z}}(\theta)$. This is related to 
the freedom that we have in choosing the default model.  
 Various  default models may be 
  allowed within the uncertainty related to $\delta$. To check this point
      we may  as well
    use the mock data that  
  causes  true flattening behavior of the free energy and study 
  the interference of such an effect with the data used in the present study.
  This will be studied in a forthcoming paper. 
 Furthermore,   some objective  analysis of the 
 parameter inference is needed with regard to the 
 regularization of the unfolding problem~\cite{rf:Cow}.  
\item As the volume increases, the flattening problem becomes more
      severe,  and
  higher  precision of  the computations becomes necessary. This is
      because a more precise calculation is required to obtain the inverse of 
      the covariance
      matrix of $P(Q)$ as the volume increases. In the case $V=8$, for
      example, the Newton method with double precision is sufficient. For
      $V=50$, however, we need quadruple precision.

\item Because the flattening behavior is inherent in the Fourier
procedure, no matter what model one treats, we have used the
Gaussian $P(Q)$  as a first attempt.
The next step is to apply the MEM to more realistic models, such as
the CP$^{N-1}$ model and  QCD, and to investigate its feasibility for them.
    The MEM analysis  might also be applicable to some other models
with the sign problem.  It may be worthwhile to study whether it is
      effective for treating theories  such as lattice field theory with
      a finite density.
\end{enumerate}

%% file: appendix1.tex
\section{SVD and the Newton Method}
\label{sec:apdx1}
\par
  In order to obtain the maximal image of ${\cal Z}(\theta)$ for fixed 
$\alpha$, we must solve the
  following equation [see Eq.~(\ref{eqn:maximumcondition})]:
\begin{equation}
  -\alpha\log\frac{{\cal Z}_{n}}{m_{n}}=\sum_{ij}K_{jn}C_{ij}^{-1}\delta P_j.
  \label{eqn:memequation}
  \end{equation}
   For later convenience, we introduce a new parameter $a_{n}$ defined
   by 
\begin{equation}
  a_{n}\equiv\log\frac{{\cal Z}_{n}}{m_{n}}. \label{eqn:newparameter}
\end{equation}\par
  We regard ${\cal Z}_{n}$ and $\bar{P}_{j}$ as $N_{\theta}$- and
  $N_{q}$-dimensional  vectors, respectively
  ($n=1,2,\cdots,N_{\theta}\;\mbox{and}\; j=1,2,\cdots,N_{q}$). In our case,
   $N_{\theta} \sim {\cal O}(10^{1})-{\cal O}(10^{2})$ and  $N_{q}\sim
  {\cal O}(10^{0})$.
   Therefore, it is non-trivial  to find
  ${\cal Z}^{(\alpha)}_{n}$ satisfying
  Eq.~(\ref{eqn:maximumcondition}). This task is made considerably
  simpler by employing the SVD. The transpose of $K$, 
$K^{t}$,  is decomposed as 
\begin{equation}
  (K^{t})_{nj}=(U\Xi V^{t})_{nj}, \label{eqn:svdforK}
\end{equation}
  where $U$ is an $N_{\theta}\times N_{q}$ matrix satisfying $U^{t}U=1$, $V$
  is an $N_{q}\times N_{q}$ orthogonal matrix,  and $\Xi$ is an
  $N_{q}\times N_{q}$ diagonal matrix,  $(\Xi)_{ij}=\xi_i
  \delta_{ij}$. The eigenvalues $\xi_i$, which are  positive
  semi-definite,  are called the
  `singular values' of $K^{t}$. They  could be arranged  by appropriate
  permutations such that
  $\xi_1\geq\xi_2\geq\cdots\geq\xi_{N_s}>\xi_{N_{s+1}}=\cdots=\xi_{N_q}=0$,
  where
\begin{equation}
  N_s={\rm rank}(K^t)\leq N_q. \label{eqn:rank}
\end{equation}
  Following Bryan\cite{rf:Bryan}, the first $N_s$ column vectors  of
  $U=({\bm u}_1,{\bm u}_2,\cdots,{\bm u}_{N_{q}})$ construct a basis in
  `singular space', which is a subspace of the $N_\theta$-dimensional space
  whose basis is   $\{{\bm u}_1,{\bm u}_2,\cdots,{\bm
  u}_{N_s}\}$, with ${\bm u}_i=(u_{1i}.u_{2i},\cdots,u_{N_\theta i})^t$.
One can then place the vector ${\bm a}$ in singular space
\begin{equation}
  {\bm a}=\sum_{i=1}^{N_{s}}\lambda_{i}{\bm u}_{i}. \label{eqn:lambdatoa}
\end{equation}
  Substituting  Eqs.~(\ref{eqn:svdforK}) and~(\ref{eqn:lambdatoa}) into
  Eq.~(\ref{eqn:memequation}), we find
\begin{equation}
  -\alpha\sum_{i}\lambda_{i}U_{ni}=\sum_{i,j}(U\Xi V^{t})_{ni}
   C_{ij}^{-1}\delta P_j. \label{eqn:memequation2}
\end{equation}
Use of  $U^{t}U=1$ then yields
  \begin{equation}
  -\alpha\lambda_{i}=\sum_{j,l}(\Xi V^{t})_{ij} C_{jl}^{-1}\delta P_l.
  \label{eqn:memequation3}
\end{equation}
  To solve $\lambda_{i}$, the  Newton method is employed.
  For each iteration, an  increment  $\delta{\bf \lambda}$ is given by
\begin{equation}
  \sum_{k}\Biggl[-\alpha\delta_{ik}-\sum_{j,l,n}(\Xi V^{t})_{ij}
   C_{jl}^{-1}  K_{ln}{\cal Z}_{n} U_{nk}\Biggr]\delta\lambda_{k} =
    \alpha\lambda_{i}+\sum_{j,l}(\Xi V^{t})_{ij} C_{jl}^{-1}\delta P_l, 
\label{eqn:newtonmethod}
\end{equation}
or in the matrix notation,
\begin{equation}
-X \delta{\bf \lambda}=\alpha\lambda+Y\delta P,
\label{eqn:newton_matrix}
\end{equation}
where $\lambda$ and $\delta P$ are $N_{q}$-dimensional vectors,   and 
$X$ and $Y$ are $N_{s}\times N_{s}$ and $N_{s}\times N_{q}$ matrices,
respectively:
\begin{equation}
X\equiv \alpha{\bf 1}+\Xi V^{t}C^{-1}K{\cal Z}U, \qquad Y\equiv \Xi 
V^{t}C^{-1}.
\end{equation}
  We then need to calculate the inverse matrix of $X$ to
  obtain $\delta{\bf \lambda}$. Note that $N_{s}=N_{q}$ holds for all the
  cases we have considered in the present paper.
  \par

%% file: appendix2.tex
\section{Derivation of Eqs.~(\ref{eqn:averageofZ2}) and
  (\ref{eqn:errorestimationforalpha}) }
  \label{sec:apdx2}
  \par
\begin{enumerate}
\item  Equation (\ref{eqn:averageofZ2}) is derived as follows.\par
\hspace*{5mm}
The probability
	${\rm prob}({\cal Z}_n|P(Q),I,m)$
         can be rewritten by use of the law of the total probability
	${\rm prob}(A)=\int dB\;{\rm prob}(A,B)$ as follows:
\begin{eqnarray}
   {\rm prob}({\cal Z}_n|P(Q),I,m)&=&\int d\alpha~
    {\rm prob}({\cal Z}_n,\alpha|P(Q),I,m) \nonumber \\
   &=&\int d\alpha~{\rm prob}({\cal Z}_n|\alpha,P(Q),I,m)
    {\rm prob}(\alpha|P(Q),I,m). \nonumber  \\ \label{eqn:totalprobability}
\end{eqnarray}
	In the second line, the definition of the conditional
	probability was  used.
	\par
         Substituting Eq.~(\ref{eqn:totalprobability}) into
         Eq.~(\ref{eqn:averageofZ}), we obtain
\begin{eqnarray}
   {\hat {\cal Z}}_n&=&\int[d{\cal Z}]{\cal Z}_n\int d\alpha~
    {\rm prob}({\cal Z}_n|\alpha,P(Q),I,m){\rm prob}(\alpha|P(Q),I,m) 
\nonumber \\
   &=&\int d\alpha~{\rm prob}(\alpha|P(Q),I,m)\int[d{\cal Z}]{\cal Z}_n~
    {\rm prob}({\cal Z}_n|\alpha,P(Q),I,m) \nonumber \\
   &\simeq&\int d\alpha~{\rm prob}(\alpha|P(Q),I,m){\cal Z}^{(\alpha)}_n,
    \label{eqn:expvalofZ}
\end{eqnarray}
       where we have assumed that the probability
	${\rm prob}({\cal Z}_n|\alpha,P(Q),I,m)$
         has a sharp peak around ${\cal Z}^{(\alpha)}_n$.
          Utilizing 
	Bayes' theorem, the law of the total probability, and the
	definition of the conditional probability, the probability
	${\rm prob}(\alpha|P(Q),I,m)$ can be rewritten as
\begin{eqnarray}
  {\rm prob}(\alpha|P(Q),I,m)&=&\frac{{\rm prob}(P(Q)|\alpha,I,m){\rm 
prob}(\alpha|I,m)}
   {{\rm prob}(P(Q)|I,m)} \nonumber \\
  &=&\frac{{\rm prob}(\alpha|I,m)}{{\rm prob}(P(Q)|I,m)}\int[d{\cal Z}]~{\rm 
prob}(P(Q),
   {\cal Z}_n|\alpha,I,m) \nonumber\\
  &=&\frac{{\rm prob}(\alpha|I,m)}{{\rm prob}(P(Q)|I,m)}\int[d{\cal Z}]~
   {\rm prob}(P(Q)|{\cal Z}_n,\alpha,I,m)   \nonumber \\
   &&\;\;\;\;\;\;\;\;\;\;\;\;\;\;\;\;\;\;\;\;\;\;\;\;\;\;\;\;\;\;\;\;
    \times {\rm prob}({\cal Z}_n|\alpha,I,m)\nonumber \\
  &\propto& {\rm prob}(\alpha|I,m)\int[d{\cal Z}]\frac{e^{W({\cal Z})}}
   {X_L X_S(\alpha)},  \label{eqn:palpha1}
\end{eqnarray}
	where Eq.~(\ref{eqn:posteriorprobability}) has been used and
       irrelevant factors, such as ${\rm prob}(P(Q)|I,m)$, have been
       ignored.\par 
\hspace*{5mm}
       Expanding $W({\cal Z})$ around $\{{\cal Z}^{(\alpha)}\}$
       up to second order, we can perform the Gaussian integration over
         configurations of $\{{\cal Z}\}$:
\begin{eqnarray}
   &&{\rm prob}(\alpha|P(Q),I,m) \nonumber \\
   &\propto& {\rm prob}(\alpha|I,m)\int[d{\cal Z}]\frac{1}
    {X_L X_S(\alpha)}
     \exp\biggl\{ W({\cal Z}^{(\alpha)})+\frac{1}{2}\sum_{n,n^\prime}
     \delta{\cal Z}_n
      \frac{\partial^2 W}
       {\partial{\cal Z}_n\partial{\cal Z}_{n^\prime}}
      \delta{\cal Z}_{n^\prime}\biggr\} \nonumber \\
   &\propto&{\rm prob}(\alpha|I,m)\exp\biggl\{\frac{1}{2}\sum_k
    \log\frac{\alpha}{\alpha +\lambda_k}+W({\cal Z}^{(\alpha)})\biggr\}
     \nonumber \\
     &\equiv& {\rm prob}(\alpha|I,m)\exp\biggl\{\Lambda(\alpha)+
      W({\cal Z}^{(\alpha)})\biggr\},
    \label{eqn:Palpha}
\end{eqnarray}
         where $\delta{\cal Z}_n\equiv{\cal Z}_n -{\cal Z}_n^{(\alpha)}$.
         Irrelevant constants have been ignored here, as before.
         The values $\lambda_k$ are eigenvalues of the real symmetric
         matrix in $\theta$ space defined in Eq.~(\ref{eqn:matrixlambda}).
         The prior probability ${\rm prob}(\alpha|I,m)$ is conventionally chosen
	to be either the Laplace rule [${\rm prob}(\alpha|I,m)=$const.] or the
	Jeffrey rule [${\rm prob}(\alpha|I,m)=1/\alpha$]. Because the integral
	in Eq.~(\ref{eqn:expvalofZ}) is insensitive to the choice of
	${\rm prob}(\alpha|I,m)$ as long as ${\rm prob}(\alpha|P(Q),I,m)$ has a 
sharp peak,
	we employ the Laplace rule
	for simplicity.~\cite{rf:AHN,rf:Bryan}\par
\hspace*{5mm}
         Substituting Eq.~(\ref{eqn:Palpha}) into
          Eq.~(\ref{eqn:expvalofZ}), we can obtain the equation for
       ${\hat {\cal Z}}_n$, 
	\begin{equation}
   {\hat {\cal Z}}_n=\frac{1}{X_W}\int d\alpha
    {\cal Z}^{(\alpha)}_n\exp\biggl\{\frac{1}{2}\sum_k
     \log\frac{\alpha}{\alpha +\lambda_k}+W({\cal Z}^{(\alpha)})\biggr\}.
\end{equation}
\vspace*{4mm}
\item Equation (\ref{eqn:errorestimationforalpha}) is derived as
       follows.\par
\hspace*{5mm}
         The uncertainty of the final output image ${\hat {\cal Z}}_n$
         is calculated as follows:~\cite{rf:JGU,rf:AHN}
\begin{equation}
   \langle (\delta {\hat {\cal Z}}_n)^2\rangle\equiv\int d\alpha
    \langle(\delta {\cal Z}^{(\alpha)}_n)^2\rangle {\rm prob}(\alpha|P(Q),I,m),
\end{equation}
         where
\begin{equation}
  \langle(\delta{\cal Z}_{m}^{(\alpha)})^2 \rangle \equiv
    \frac{\int[d{\cal Z}]\int_\Theta d\theta_n d\theta_{n^\prime}
     \delta{\cal Z}_n\delta{\cal Z}_{n^\prime} {\rm prob}({\cal 
Z}_m|P(Q),I,m,\alpha)}
      {\int[d{\cal Z}]\int_\Theta d\theta_n d\theta_{n^\prime}
       {\rm prob}({\cal Z}_m|P(Q),I,m,\alpha)}.
\end{equation}
Using $\int[d{\cal Z}]~{\rm prob}({\cal Z}_m|P(Q),I,m,\alpha)\equiv 1$ and
inserting Eq.~(\ref{eqn:posteriorprobability}), we obtain
\begin{equation}
   \langle(\delta{\cal Z}_{m}^{(\alpha)})^2 \rangle
   =\frac{1}{\int_\Theta d\theta_n d\theta_{n^\prime}}
    \int_\Theta d\theta_n d\theta_{n^\prime}\int[d{\cal Z}]
     \delta{\cal Z}_n\delta{\cal Z}_{n^\prime}
      \frac{e^{W({\cal Z})}}{X_L X_S(\alpha)}.
\end{equation}
When $e^{W({\cal Z})}$ is
         expanded around $\{{\cal Z}^{(\alpha)}\}$ up to second order,
         and the integral over ${\cal Z}_n$ is performed,
         Eq.~(\ref{eqn:errorestimationforalpha}) is derived as
\begin{eqnarray}
  \langle(\delta{\cal Z}_{m}^{(\alpha)})^2 \rangle
  &=&\frac{1}{\int_\Theta d\theta_n d\theta_{n^\prime}}
   \int_\Theta d\theta_n d\theta_{n^\prime}\int[d{\cal Z}]
    \delta{\cal Z}_n\delta{\cal Z}_{n^\prime}\frac{1}{X_L X_S(\alpha)} 
\nonumber \\
  &&\;\;\;\;\;\;\;\;\;\;\;\;\;\;\;\;\;\times
    \exp\biggl\{ W({\cal Z}^{(\alpha)})+\frac{1}{2}\sum_{n,n^\prime}
    \delta{\cal Z}_n\frac{\partial^2 W}
    {\partial{\cal Z}_n\partial{\cal Z}_{n^\prime}}
      \delta{\cal Z}_{n^\prime}\biggr\} \nonumber \\
   &\simeq&- \frac{1}{\int_\Theta d\theta_n d\theta_{n^\prime}}
    \int_\Theta d\theta_n d\theta_{n^\prime} \Bigl(\frac{\partial^2 W}
     {\partial{\cal Z}_n\partial{\cal Z}_{n^\prime}}
      \Bigm|_{{\cal Z}={\cal Z}^{(\alpha)}}\Bigr)^{-1}.
\end{eqnarray}
\end{enumerate}

%% file: appendix3.tex
\section{Uniqueness of the Maximum of $W$}
  \label{sec:apdx3}
  \par
   Following Asakawa et. al.~\cite{rf:AHN}, let us check  that a unique
  maximum of $W$  exists for $\alpha\neq 0 ~
(>0)$.  In our case, the kernel is given by that of  the Fourier transform.
   The curvature of $W$ is given by
  \begin{equation}
\frac{\partial^2 W}{\partial {\cal Z}_m \partial {\cal Z}_n}=-\alpha
\delta_{mn}\frac{1}{{\cal Z}_n}-\sum_{ij}K_{im}C^{-1}_{ij}K_{jn}.
  \end{equation}
  Introducing   any $N_\theta$-dimensional real vector ${\bm y} (\neq 0)$,
let us calculate
    \begin{equation}
\sum_{mn}y_m\frac{\partial^2 W}{\partial {\cal Z}_m \partial
{\cal Z}_n}y_n=\sum_{mn}y_m\Biggl[-\alpha
\delta_{mn}\frac{1}{{\cal Z}_n}-\sum_{ij}K_{im}C^{-1}_{ij}K_{jn}\Biggr] y_n.
\label{curvW}
  \end{equation}
  \begin{enumerate}
  \item $\alpha=0$ case\par
  When $\alpha=0$, Eq.~(\ref{curvW}) becomes
   \begin{equation}
\sum_{mn}y_m\frac{\partial^2 W}{\partial {\cal Z}_m \partial
{\cal Z}_n}y_n=-\sum_{mn}y_m\sum_{ij}K_{im}C^{-1}_{ij}K_{jn}y_n.
\label{curv}
  \end{equation}
  Because the covariance matrix $C$ is symmetric, $C$ is diagonalized by  an
$N_q\times N_q$ orthogonal matrix $R$ as
  $$R^t C R = \bar {\sigma}^2. $$
  Also, defining the $N_q\times N_\theta$ matrix
  $$\tilde K\equiv R^tK $$
  and the $N_q$-dimensional real vector
  \begin{equation}
  \tilde {\bm y} \equiv \tilde K {\bm y},
  \label{ytild}
  \end{equation}
  Eq.~(\ref{curv}) becomes
   \begin{equation}
\sum_{mn}y_m\frac{\partial^2 W}{\partial {\cal Z}_m \partial
{\cal Z}_n}y_n= -\sum_{i}\frac{\tilde y^2_i}{\bar
{\sigma}^2_i }\leq 0.
  \end{equation}
  This could vanish only for $\tilde {\bm y}=0$,  and this is 
realized for
  non-trivial vectors ${\bm y}$, because Eq.~(\ref{ytild}) asserts that
the dimension of the solution vector space
  is
  $$N_\theta - {\rm rank} \tilde K \geq N_\theta- N_q  > 0 .$$
  In the $\alpha=0$ case, therefore, there are multiple maxima of $W$.
  \item $\alpha\neq 0$ case\par
  For  $\alpha\neq 0$,  Eq.~(\ref{curvW}) becomes
    \begin{equation}
\sum_{mn}y_m\frac{\partial^2 W}{\partial {\cal Z}_m \partial
{\cal Z}_n}y_n=-\alpha\sum_{n} \frac{y_n^2}{{\cal Z}_n}-\sum_{i}\frac{\tilde
y^2_i}{\bar {\sigma}^2_i}.
  \end{equation}
  Because ${\cal Z}_n > 0$,  the curvature of $W$  becomes negative definite:
  $$\sum_{nm}y_m\frac{\partial^2 W}{\partial {\cal Z}_m \partial {\cal 
Z}_n}y_n<0.$$
  Therefore, the entropy term is essential to the uniqueness of the
maximum.
  \end{enumerate}

%% file: appendix4.tex
   \section{Comparison with the Fourier Method}
    \label{sec:apdx4}
    \par
In the  case of \underline{no flattening}, we show
   that the analysis  in the Newton method leads
to the same result as that obtained using the Fourier method.
This holds  for   $\alpha=0$  and $\alpha \neq 0 $ (but only small $\alpha>0$).
Note  that in the Fourier method,
  \begin{equation}
  {\bar P}=K{\cal Z}
  \label{Four}
   \end{equation}
  is successfully inverted  in the case without flattening.
  This is because $P(Q)$ is a rapidly decreasing function,  and ${\cal
  Z}(\theta)$, which is given by the Fourier transform,
  is  smooth enough   that the contribution from
  higher $Q$ can be ignored.  \par
  \vspace*{4mm}
    \begin{enumerate}
  \item $\alpha=0$ case\par
  Let us first consider the $\alpha=0$ case.
  For $\alpha=0$, Eq.~(\ref{eqn:newton_matrix})   reduces to
  \begin{equation}
-X _0 \delta{\bf \lambda}=Y\delta P,
\label{eqn:newton_matrix0}
\end{equation}
where
  \begin{equation}
  X_0 \equiv\Xi V^tC^{-1}KZU.
  \label{defX}
  \end{equation}
  When $X_0$ is regular, the increment $\delta\lambda$ becomes
  \begin{equation}
  \delta\lambda =-X_0 ^{-1}Y\delta P.
  \end{equation}
  When  the integrations converge, i.e., when $\delta\lambda =0$, in the
	Newton method, 
  we find
  \begin{equation}
  X_0^{-1}Y\delta P=0.
  \end{equation}
Because the $N_q\times N_q$ matrix  $ X_0^{-1}Y$ is regular (this can be 
checked numerically),
\begin{equation}
  \delta P=K{\cal Z}-\bar{P}=0
  \label{eqn:fixed}
  \end{equation}
is a unique solution. \par
\hspace*{5mm}
Equation (\ref{eqn:fixed})  gives $N_q$ equations:
\begin{equation}
\sum_{n=1}^{N_\theta}K_{in }m_n\exp(\sum_{j=1}^{N_q} U_{nj}\lambda_j) =\bar 
{P}_i
\qquad   ( i=1,\dots , N_q  ) .
\label{Pi2}
\end{equation}
Thus, one obtains  a unique solution for $\lambda_i
(i=1,\dots, N_q)$ for a given default $m_n$.
  This  turns  out to  give a 
${\cal Z}_n$ that is equivalent to  that found in the Fourier method.
\vspace*{4mm}
\item $\alpha\neq 0$ case (small $\alpha$)\par
When $\alpha\neq 0$, we use Eq.~(\ref{eqn:newton_matrix})  in the Newton 
method.
  For small $\alpha$, this is approximated by
  \begin{equation}
  X\delta\lambda =- Y \delta P.
  \end{equation}
  By using  the regularity of  $X$, $\delta\lambda$ is given by
  \begin{equation}
  \delta\lambda =-X^{-1}Y\delta P.
  \end{equation}
  When the integration converges, $\delta \lambda=0$, the regularity of
	$X^{-1}Y$ implies 
  \begin{equation}
  \delta P=0,
  \end{equation}
  which is the same as Eq.~(\ref{eqn:fixed}).
  Therefore we obtain a unique solution in this case too.
\end{enumerate}
\vspace*{4mm}
We thus find that for $\alpha=0$ and/or for small values of $\alpha$,
  the Newton method leads
to the same result  ${\cal Z}^{(\alpha)}_n$ as that obtained using 
the Fourier method.
Moreover, if the  probability ${\rm prob}(\alpha|P(Q),I,m)$ dominates
  for $\alpha\approx 0$,
  then the averaged image $ {\hat {\cal Z}}_n$ is also in good agreement
  with that obtained from the Fourier method. Actually, this is true for the
  cases without flattening,   as discussed in
  \S~\ref{sec:result_noflat}. \par